\documentclass[10pt]{article}
\usepackage{amsmath, amsfonts}
\usepackage{mathrsfs,amssymb}
\usepackage{graphicx}
\usepackage{lscape}
\usepackage{hyperref}
\usepackage{bbm}
\usepackage{rotating}
\usepackage{float}
\numberwithin{equation}{section}
\usepackage{multirow}
\usepackage{setspace}
\usepackage{xcolor}
\usepackage{subcaption}
\usepackage{authblk}
\providecommand{\keywords}[1]{\textbf{\textit{keywords---}} #1}

\begin{document}

\title{Deciphering hierarchical organization of topologically associated domains through change-point testing}
\author[1]{Haipeng Xing\footnote{Haipeng Xing and Yingru Wu have contributed equally}}
\newcommand\CoAuthorMark{\footnotemark[\arabic{footnote}]} 
\author[1]{Yingru Wu\protect\CoAuthorMark}
\affil[1]{Department of Applied Mathematics and Statistics, State University of New York at Stony Brook}
\author[2]{Michael Q. Zhang}
\affil[2]{Center for System Biology, University of Texas at Dallas}
\author[3]{Yong Chen\footnote{Correspondence: chenyong@rowan.edu}}
\affil[3]{Department of Molecular and Cellular Biosciences, Rowan University}

\date{}
\maketitle

\begin{abstract} 
\textbf{Background} The nucleus of eukaryotic cells spatially packages chromosomes into a hierarchical and distinct segregation that plays critical roles in maintaining transcription regulation. High-throughput methods of chromosome conformation capture, such as Hi-C, have revealed topologically associating domains (TADs) that are defined by biased chromatin interactions within them. \newline
\textbf{Results} Here, we introduce a novel method, HiCKey, to decipher hierarchical TAD structures in Hi-C data and compare them across samples. We first derive a generalized likelihood-ratio (GLR) test for detecting change-points in an interaction matrix that follows a negative binomial distribution or general mixture distribution. We then employ several optimal search strategies to decipher hierarchical TADs with p-values calculated by the GLR test. Large-scale validations of simulation data show that HiCKey has good precision in recalling known TADs and is robust against random collision noise of chromatin interactions. By applying HiCKey to Hi-C data of seven human cell lines, we identified multiple layers of TAD organization among them, but the vast majority had no more than four layers. In particular, we found that TAD boundaries are significantly enriched in active chromosomal regions compared to repressed regions, indicating finer hierarchical architectures in active regions for precise gene transcription regulation. \newline
\textbf{Conclusions} HiCKey is optimized for processing large matrices constructed from high-resolution Hi-C experiments. The method and theoretical result of the GLR test provide a general framework for significance testing of similar experimental chromatin interaction data that may not fully follow negative binomial distributions but rather more general mixture distributions.
\end{abstract}


\section*{Background}

The eukaryotic genome is hierarchically organized in the nucleus, exhibiting well-maintained three-dimensional (3D) structures for its cellular functions. DNA and associated proteins constitute chromatin units, among which interactions are not random but precisely regulate transcription and replication during the cell cycle \cite{Cavalli:2013,Gibcus:2013, Bonev:2016}. For example, the interactions between enhancers and their distal targeted genes are essential for controlling gene expression strengths and tissue-specific expression patterns \cite{new21}. 3D chromosomal studies in prostate cancer, thalassemia, breast cancer and multiple myeloma have revealed that disordered interactions are closely related to gene dysregulation, contributing to the development of cancer and other genetic diseases \cite{new22, new23, Spielmann:2018}. Thus, estimating the 3D organization of chromosomes can provide important insight into not only the role of high-order chromatin compaction in gene regulation but also the way disordered chromatin interactions lead to diseases.

To systematically delineate chromatin interactions and 3D organization, novel experimental methods have been developed by employing high-throughput sequencing techniques. Chromosome conformation capture (3C) \cite{Dekker:2002} and its high-throughput derivatives, such as ChIA-PET \cite{Tang:2015}, HiChIP \cite{new24} and Hi-C \cite{Dixon:2012, Rao:2015}, have granted researchers comprehensive information on chromatin interactions and hierarchical chromosomal organizations, including active or repressive compartments (A/B compartments) \cite{Naumova:2010}, topologically associated domains (TADs) \cite{Dixon:2012, Crane:2015, Nora:2012}, CTCF protein-mediated loops \cite{Tang:2015} and enhancer-promoter interactions \cite{new21}. In general, a chromosome can be divided into active or repressed compartments (A/B compartments) corresponding to higher or lower gene expression levels \cite{Naumova:2010}. The analysis of high-resolution Hi-C data has shown that chromosomes can be divided into functional units, called TADs, which are conserved across multiple human and mouse cell lines \cite{Dixon:2012, Dixon:2016, Rao:2015}. Furthermore, ChIA-PET data of CTCF, Cohesin and RNA PolII have revealed fine spatial structures of CTCF loops and enhancer-promoter interactions \cite{Tang:2015, new21}. Compared with ChIA-PET and other capture-based methods, Hi-C provides high-resolution unbiased signals of chromatin interactions \cite{Rao:2015}.

\begin{figure}[h!]
\caption{HiCKey workflow.}
\includegraphics[width=0.96\textwidth]{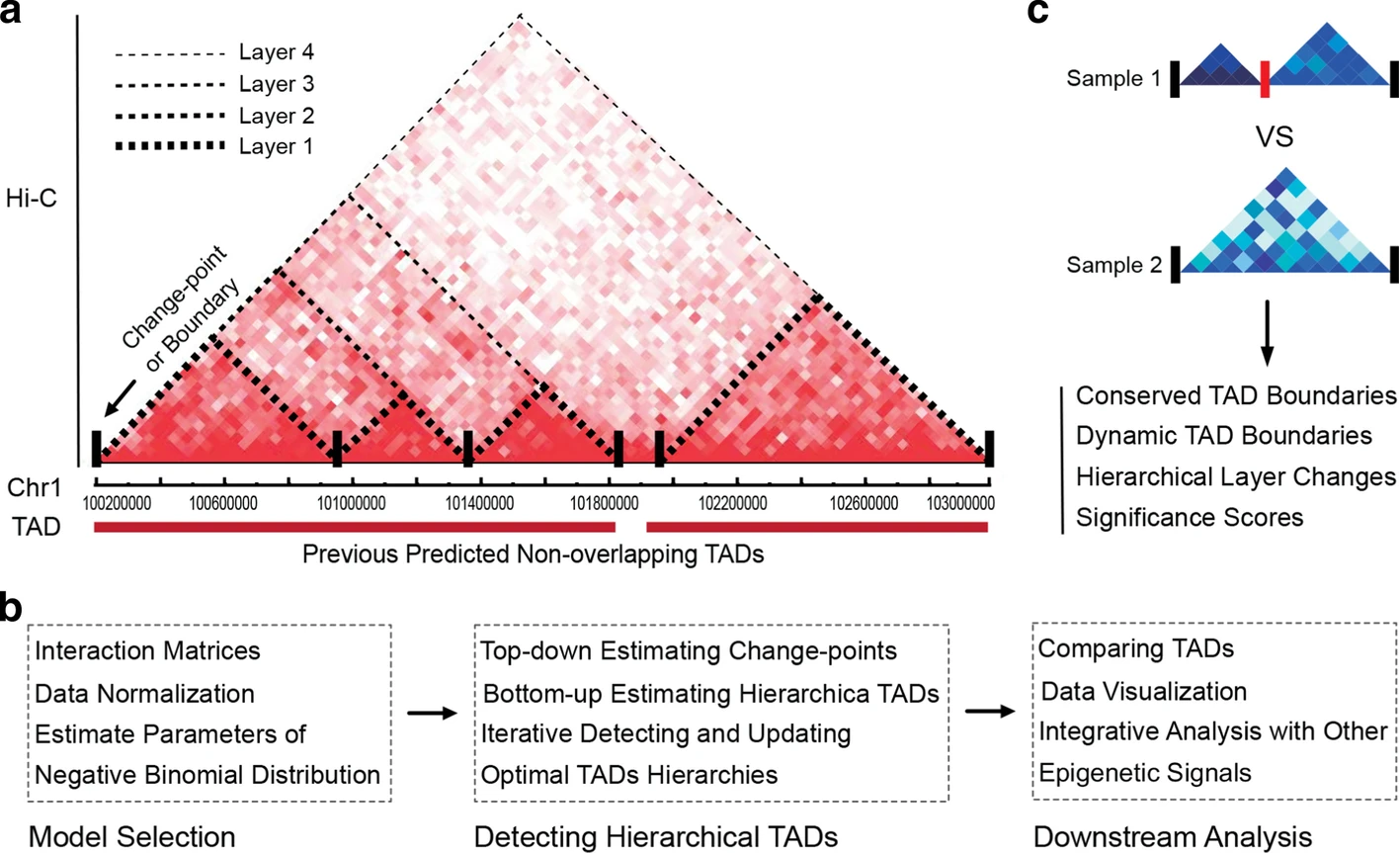}\par
\textbf{a.} Illustration of hierarchical organizations of chr1:100Mb-103Mb from \emph{in situ} Hi-C data of GM12878. The potential four-layer organization is denoted by dotted lines. The non-overlapping TADs were obtained from a previous study \cite{Rao:2015}. \textbf{b.} The workflow of HiCKey. HiCKey takes interaction matrices (or interaction list) as input, iteratively searches TADs, and outputs their hierarchies. It can also provide additional steps for TAD comparison and data visualization. \textbf{c.} Comparing two sets of TADs structures.
\label{Fig1_Workflow}
\end{figure}

TADs can be considered isolated structures that partition chromosomes into discrete functional regions and thus restrict regulatory activities within them \cite{Dixon:2012, Crane:2015, Bonev:2016}. To detect TAD structures from Hi-C data, many computational methods have been proposed by calculating the insulation scores or defining significance values of TAD boundaries. However, most of them are tools for detecting nonhierarchical TADs, such as Armatus \cite{Filippova:2014}, TopDom \cite{Shin:2015}, HiCSeg \cite{Levy-leduc:2014}, InsulationScore \cite{Gong:2017}, Arrowhead \cite{Rao:2015} and DomainCaller \cite{Dixon:2012}. Since TADs were shown to be hierarchically organized \cite{Dixon:2012, Rao:2015}, the estimated nonhierarchical TADs cannot fully describe the biological hierarchy in cell systems. As shown in Fig. \ref{Fig1_Workflow}a, a $\sim$3 Mb region on chr1 of the GM12878 cell line clearly exhibits four layers of TADs with different interaction strengths. To overcome the limitations of nonhierarchical TAD finders, another type of method, such as TADtree \cite{Weinerb:2015}, IC-Finder \cite{Haddad:2017}, GMAP \cite{Yu:2017}, Matryoshka \cite{Malik:2018} and 3DNetMod \cite{3dnetmod}, has been proposed to find TADs and their nested sub-TAD organizations. Although these methods have given researchers new knowledge in understanding chromosomal organization, they still suffer from low precision or poor robustness against noise or high time consumption \cite{Marie:2018}. IC-Finder \cite{Haddad:2017} employs a constrained hierarchical clustering strategy that iteratively groups objects into a hierarchy of clusters. Although it was robust against noise, it requires high sequencing depth \cite{Marie:2018}. Another method, GMAP \cite{Yu:2017}, utilizes a Gaussian mixture model to iteratively identify TADs but is limited to two levels of TADs. TADtree \cite{Weinerb:2015} finds the best TAD hierarchy via a dynamic programming algorithm that was tested to be time consuming for large size Hi-C matrices \cite{Marie:2018}. 3DNetMod \cite{3dnetmod} uses network modularity theory to hierarchically cluster TADs; however, it is sensitive to multiple parameter settings and is less robust against experimental noise. One major challenge in detecting TAD structures is the experimental noise that mainly comes from random ligation of chromosomal segments during the cross-linking step and the “genomic distance effect” in Hi-C experiments, reducing the consistency and prevalence of higher-order structures \cite{Dixon:2012, Rao:2015, nobel}. Another obstacle in Hi-C data analysis is how interaction frequencies are distributed. Negative binomial (NB) distribution is the most widely used assumption, but it cannot fully capture the characteristics of chromatin interactions \cite{Dixon:2016, Rao:2015} since confounding factors of Hi-C experiments may transform the interaction frequencies into more complicated distributions (e.g., a mixture of unknown discrete distributions). Thus, there are urgent requirements for new TAD detection methods to precisely estimate chromosome structure.

Understanding dynamic changes in TADs is also an important topic in Hi-C data analysis since disordered TADs are linked with cell-specific gene expression regulation or different developmental conditions. For example, Sauerwald and Kingsford et al. confirmed that conservation and dynamics of TAD boundaries were associated with distinct biological conditions or chromosomal variations by comparing a large number of Hi-C experiments of cell lines or tissues \cite{Sauerwald:2018, Sauerwald:2020}. Several methods have been proposed for detecting boundary changes in TADs, including HiCcompare \cite{new19}, localTADSim \cite{Sauerwald:2020}, HOMER \cite{HOMER:2018}, HiCDB \cite{Fengling:2018} and TADCompare \cite{Cresswell:2020}. The major strategy of these methods is to first detect TADs separately and then compare two sets of TAD boundaries. However, these methods usually require specific data types and lack statistical rigorousness. HOMER \cite{HOMER:2018} only outputs different TAD regions by overlapping two sets of TADs but does not provide significance testing for boundary differences. Another method, TADCompare \cite{Cresswell:2020}, has arisen as a potentially useful tool for comparing TAD boundaries. This method proposes a new boundary score for differential boundary detection, time-course analysis of boundary changes, and consensus boundary calling but is limited to five types of boundary changes. LocalTADSim \cite{Sauerwald:2020} requires using Armatus software or manually formatting their inputs as Armatus output. HiCDB \cite{Fengling:2018} uses a new metric named relative local insulation that is similar to insulation score, but it is biased to top-ranked insulation scores.

Based on the above observations, we propose a novel computational method, called HiCKey, to decipher the hierarchical organization of chromatin interactions in Hi-C data (Fig. \ref{Fig1_Workflow}b). 
We derived a generalized likelihood-ratio test (GLR) for calling TAD boundaries (change-points), which is a matrix-variant change-point testing method in the literature. HiCKey can be applied to different interaction strength distributions. This is important for statistical analysis of Hi-C data, which is composed of biological interactions, random missing interactions and random ligation noise. Furthermore, the p-values of a change-point from different Hi-C matrices can be combined by Fisher's method, providing a measure of whether a boundary is conserved across different samples (Fig. \ref{Fig1_Workflow}c). We demonstrated the performance and robustness of HiCKey using substantial validations of simulation studies. By applying HiCKey to seven human cell lines, we identified not only multiple layers of TAD organization in each cell line but also TAD structures consisting of different gene expression or histone modification signals. We found that TAD boundaries are significantly enriched in active chromosomal regions, indicating that fine TAD architectures are employed for precise gene transcription control. These results show the advantages of HiCKey in detecting TADs and provide novel biological discoveries revealing the association of chromosomal organization and gene regulation.

\section*{Methods}

\subsection*{Modelling TADs organization}

\begin{figure}[h!]
\caption{Schematic plot of hierarchical TAD detection.}
\includegraphics[width=0.96\textwidth]{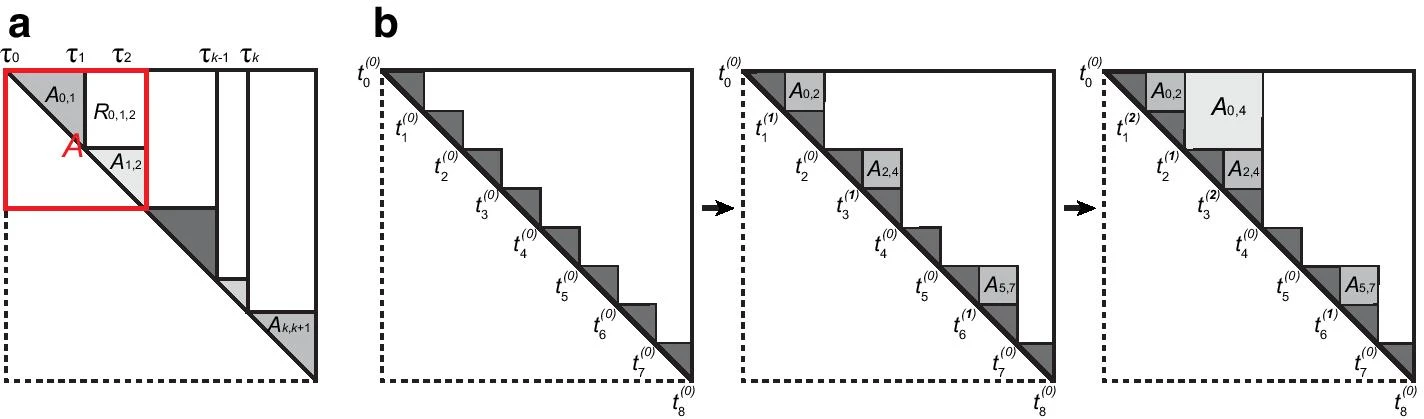}\par
\textbf{a.} Block diagonal with $K$ change-points and one change-point (red region). Sub-matrices are shown with a red region that has a single change-point. \textbf{b.} The bottom-up procedure for detecting hierarchical TADs. At each iteration, the p-values are iteratively calculated for all boundaries, and the layer labels of inner boundaries of two blocks are iteratively updated if they satisfy merge conditions.
\label{Fig2_Schematic}
\end{figure}

Hi-C experiments generate a symmetric $n$ by $n$ matrix ${\bf X}=(x_{ij})\in \mathbbm{R}^{n\times n}$, where $n$ is the number of bins and $x_{ij}$ is the frequency of chromatin interactions between a pair of genomic loci $i$ and $j$. Assume there are $K$ change-points located at $1 = \tau_0 < \tau_1 < \tau_2< \dots < \tau_K < n$. These change-points divide all chromosomal bins into $K+1$ non-overlapping TADs, as shown in Fig. \ref{Fig2_Schematic}a. For change-points $\tau_a$, $\tau_b$ and $\tau_c$, the TAD between $\tau_a$ and $\tau_b$ is $A_{a,b}$. The rectangle between $A_{a,b}$ and $A_{b,c}$ is $R_{a,b,c}$ ($R$). We aim to detect (1) all the change-points and (2) the hierarchical organization of these change-points.

\noindent

Previous Hi-C studies have revealed that within-TAD interactions are much stronger than cross-TAD interactions \cite{Dixon:2012, Dixon:2016, Rao:2015}, as shown in a Hi-C matrix of GM12878 (Fig. \ref{Fig1_Workflow}a). In general, the average interactions in neighbouring blocks, $A_{0,1}$ and $A_{1,2}$, can be different, but they are both stronger than the interactions in $R$ (Fig. \ref{Fig2_Schematic}a). Additionally, the interaction strength decreases as the distance from the diagonal increases. These biological observations raise statistical insight that a change-point, $\tau_1$, will lead to significant distribution differences among $A_{0,1}$, $A_{1,2}$ and $R$.

In theoretical statistics, this problem is known as change-point analysis. We note that various frequentist and Bayesian methods have been developed for multiple change-point analyses of uni- and multi-variate data over the last few decades. In particular, Bayesian methods assume that multiple change-points follow a stochastic process and solve the inference problem through Markov chain Monte Carlo (MCMC) simulations \cite{Liu:1999, Wang:2000}. Exact Bayesian inference approaches with efficient approximations were also developed for various problems \cite{Lai:2013, Xing:2012a, Xing:2012b}. The frequentist methods provide more technical tools, including dynamic programming algorithms to solve maximum likelihood estimation \cite{Bai:1998, Perron:2007}, binary segmentation \cite{Matteson:2014}, information criteria model selection \cite{Jshen:2012, Zhang:2007}, and penalized likelihood or cost function approaches \cite{Lavielle:2005, Harchaoui:2010}. However, none of these methods can be directly applied to Hi-C data because they are matrix-variate. Simplifying them to multi-variate vectors and feeding them to existing change-point methods are not optimal. Furthermore, a general tool needs to be developed to address cases that are more complicated than some single distribution assumption (e.g., NB).

\subsection*{Generalized likelihood-ratio test for change-points}

In Fig. \ref{Fig2_Schematic}a, the red block is a sub-matrix from $\tau_0$ to $\tau_2$ in matrix $X$. Its upper triangular part is denoted by $A$. Assume that all interaction reads are independent random variables from an NB family. If there exists a change-point, we then have three sets of block-wise constant parameters. Otherwise, all the parameters will be the same. Specifically,
\begin{align*}
x_{ij} \sim NB(\mu_k, r), \quad 1\le i \le j \le n, \quad
\mu_k = \left\{ \begin{array}{ll}
\mu_1, & \mbox{if} (i,j) \in A_{0,1}\\
\mu_2, & \mbox{if} (i,j) \in A_{1,2}\\
\mu_0, & \mbox{if} (i,j) \in R
\end{array} \right.
\end{align*}
where $\mu_k$ is the mean of the NB distribution and $r$ is a nuisance parameter with a positive value. In $A$, we consider the hypothesis test $H_0$: there is no change-point against $H_1$: there is one change-point at unknown position $\tau_1 = m (1 < m < n)$, such that $A=A_{0,1}\cup A_{1,2} \cup R$. Let $S_A$, $S_{A_k}$ and $S_R$ be the sums of $x_{ij}$ in the corresponding regions.
\begin{align*}
\resizebox{0.9\linewidth}{!}{$S_A=\sum_{(i,j) \in A} x_{ij}, \quad S_R = \sum_{(i,j) \in R_{0,1,2}} x_{ij}, 
\quad S_{A_1} = \sum_{(i,j) \in A_{0,1}} x_{ij}, \quad S_{A_2} = \sum_{(i,j) \in A_{1,2}} x_{ij}$}
\end{align*}
\noindent When the location of change-point $\tau_1 = m$ is known, the logarithm of the generalized likelihood-ratio test statistic is
\begin{align*}
GLR_{NB,m} &= \sum_{k=1,2} \Big\{ S_{A_k} \log \Big(
\frac{S_{A_k}/|A_{k-1,k}|}{r+S_{A_k}/|A_{k-1,k}|} \Big)\\
&+ r |A_{k-1,k}| \log \Big( \frac{r}{r+S_{A_k}/|A_{k-1,k}|} \Big) \Big\}\\
&+ \Big( S_R \log \Big( \frac{S_R/|R|}{r+S_R/|R|} \Big)
+ r |R| \log \Big( \frac{r}{r+S_R/|R|} \Big) \Big)\\
&- \Big( S_{A} \log \Big( \frac{S_{A}/|A|}{r+S_{A}/|A|} \Big)
+ r |A| \log \Big( \frac{r}{r+S_{A}/|A|} \Big) \Big)
\end{align*}
\noindent where $| \cdot |$ is the cardinality of a set.

\noindent {\bf Theorem 1}. {\it The GLR statistic, $GLR_{NB,m}$, is asymptotically equivalent to the following scan statistic if $m/n$ holds constant as $n \rightarrow \infty$
\begin{equation}\label{glr2}
\resizebox{.9\hsize}{!}{$
Z_m := \frac{1}{2\sigma_0^2}\bigg\{ \frac{(S_{A_1}-\frac{|A_{0,1}|}{|A_{0,1}\cup R|} S_{A_1
\cup R})^2}{|A_{0,1}| (1-\frac{|A_{0,1}|}{|A_{0,1}\cup R|})} +
\frac{(S_{A_1\cup R}-\frac{|A_{0,1} \cup R|}{|A|}
S_{A})^2}{|A_{0,1}\cup R| (1-\frac{|A_{0,1}\cup R|}{|A|})} \bigg\}
$}
\end{equation}
\noindent where $S_{A_1 \cup R} = S_{A_1} + S_R$, $\sigma_0^2$ is the variance under the null hypothesis, which can be estimated.}

The first term on the right-hand side in \eqref{glr2} describes the difference between $A_{0,1}$ and $R$, while the second term describes the difference between $A_{0,1} \cup R$ and $A_{1,2}$. Note that if the change-point position is known, the partition ratio $m/n$ of the matrix $A$ is fixed. Hence, it is natural to assume that m/n holds constant as $n \rightarrow \infty$. It is easy to see that the asymptotic distribution of $Z_m$ is chi-square. However, in practice, $m$ is unknown, so we need to define a new test statistic and study its asymptotic properties.

\noindent {\bf Definition 1}. {\it We define the following test statistic:
\begin{equation}\label{statZ}
\widetilde{Z}=\max_{\xi < m \le n-\xi} Z_m,
\end{equation}
\noindent $Z_m$ is calculated for each $(\xi< m \le n-\xi)$, and we take the supremum, where $\xi$ is the minimum size of a TAD.
}

The minimal TAD size is defined by default as the up integer of 100 kb/resolution. Here, 100 kb is estimated by the observed sub-TAD size in real biological datasets, since more than 95\% of sub-TAD sizes are larger than 100 kb among multiple cell types of Rao’s Hi-C data \cite{Rao:2015}. For example, for the interaction matrix with 40k resolution, 3 (round 100/40 up to 3) is set as the minimal TAD size. Moreover, the p-value threshold and the minimal TAD size threshold are flexible in HiCKey software and can be changed by users for specific research goals.

As a result, we can eliminate the assumption on the original distribution. The above $\widetilde{Z}$ can be used to detect the change in means and is not limited to an NB distribution. Under the null hypothesis, we take the upper triangular part, $A$, as a discrete-time random-walk with a two-dimensional time index. By Donsker's invariance principle, we obtain the asymptotic distributions of $Z_m$ and $\widetilde{Z}$ in the following theorem.

\noindent {\bf Theorem 2} {\it Consider a Gaussian random field $G(s, t)$, with location indices $s$ and $t$, defined on the upper triangular part of a unit square $B=\{ (s,t) | 0\le s \le t \le 1 \}$. Assuming that $m/n \rightarrow t \in (0, 1)$ as $n \rightarrow \infty$, then the regions $\frac{A_{0,1}}{\sqrt{n^2/2}}, \frac{A_{1,2}}{\sqrt{n^2/2}}$ and $\frac{R}{\sqrt{n^2/2}}$ converge to regions $\widetilde{A}_1= \{ (\widetilde{s}, \widetilde{t}) | 0 \le \widetilde{s} \le \widetilde{t} \le t \}$, $\widetilde{A}_2=\{ (\widetilde{s}, \widetilde{t}) | t \le \widetilde{s} \le \widetilde{t} \le 1 \}$, and $\widetilde{R}=B - \widetilde{A}_1 - \widetilde{A}_2$, respectively (note that $\frac{A}{\sqrt{n^2/2}} \rightarrow B$). Correspondingly,
\begin{equation*}
Z_m \rightarrow g_t := \frac{(G_{\widetilde{A}_1}-\frac{t}{2-t} G_{\widetilde{A}_1
\cup \widetilde{R}})^2}{2t^2(1-t)/(2-t)} +
\frac{(G_{\widetilde{A}_1\cup \widetilde{R}}-t(2-t)
G_{\widetilde{A}})^2}{t(1-t)^2(2-t)},
\end{equation*}
\noindent and if $\xi/n \rightarrow \delta >0$,
\begin{equation*}
\widetilde{Z} \rightarrow g_\delta := \max_{\delta < t <1-\delta}
g_t.
\end{equation*}
}
Details of the Gaussian random field construction and proof are included in Additional file 1. This asymptotic property is extremely helpful in high-resolution Hi-C data. Consider a TAD with a fixed chromosomal size (typically 1 Mb), where the higher the resolution is, the more reads TAD contains in the Hi-C matrix. In practice, Monte Carlo simulations can be used to obtain the asymptotic distribution from the above theorem. A histogram and a kernel density estimation are included in Additional file 1, Fig. S1. Since our GLR testing theoretically converges for different types of distributions, the parameters can be applied to different datasets.

\subsection*{Detecting hierarchical TADs}

We propose an iterative algorithm to implement the GLR test in estimating hierarchical TADs. The first step is binary segmentation to identify all the change-points (TAD boundaries). In the Hi-C matrix, we first find one change-point that has the maximum $Z_m$ in Eq (\ref{statZ}), resulting in two diagonal sub-matrices. Iteratively, one change-point is found for each sub-matrix, until the sub-matrix has a size smaller than the lower bound, $2\xi$. In the second step, we use a pruning process to test each change-point in reverse order to which they are identified and to remove insignificant change-points. A $p$-value threshold, $\alpha_0$, is needed for all the tests.

\medskip
\hrule
\medskip
\noindent (binary segmentation) \newline
\noindent \textbf{while} there is a diagonal block, $A$, with size $T \geq 2\xi$\newline
\indent \indent \textbf{for} $i$ from $\xi + 1$ to $T - \xi + 1$ in $A$ \newline
\indent \indent \indent \indent calculate $Z_i$ \newline
\indent \indent find a change-point at $\arg\max Z_i$ \newline
\indent \indent record the order we identify that change-point \newline
\noindent (pruning) \newline
\noindent \textbf{for} each change-point, $\tau_t$, in reverse order to which they are identified \newline
\indent \indent take the sub-matrix from its left closest $\tau_{t - 1}$ to right closest $\tau_{t + 1}$ and \newline
\indent \indent calculate the GLR test statistic $\widetilde{Z} = Z_t$ \newline
\indent \indent \textbf{If} $p$-value $> \alpha_0$ \newline
\indent \indent \indent eliminate the change-point
\medskip
\hrule
\medskip

Although binary segmentation in a top-down strategy can provide hierarchical organization as divisive clustering or a decision tree, it may contain more false hierarchical structures. We use a bottom-up procedure to merge neighbouring blocks and update the layer labels of boundaries (please see Fig. \ref{Fig2_Schematic}b for a demo example). More specifically, we recalculate the p-value for each potential boundary outputted in the top-down step by testing its flanking blocks with the attached rectangle sub-matrix. The p-values of all boundaries are ranked in descending order, and their layer labels are initialized as zero. In descending order, two neighbouring blocks are parallelly merged into one if their inner boundary p-value is larger than a threshold $\alpha_1=1e-5$. The layer label of the inner boundary of merged blocks increases by one. In each iteration, the boundary p-values between merged blocks are recalculated. The iteration continues until no remaining blocks satisfy the merge conditions. Here, $\alpha_0$ and $\alpha_1$ are used to control the number of potential TAD boundaries and the number of hierarchical branches, respectively. In HiCKey, $\alpha_1=1e-5$ was used by default, which can well-delineate local hierarchical structures in real data analysis. These two parameters can be reset by users for different TAD detection goals. If $\alpha_1$ increases to $\alpha_0$, more blocks are considered as individual TADs (less hierarchical). In contrast, if $\alpha_1$ is smaller, more blocks are grouped into hierarchical structures.

\medskip
\hrule
\medskip

\noindent Recalculate the p-values for all boundaries by using two neighbouring blocks and their attached rectangle sub-matrix. \newline
\noindent \textbf{While} there are p-values of boundaries larger than $\alpha_1=1e-5$. \newline
\indent \textbf{for} each boundary $t_m$ in descending order\newline
\indent \indent \textbf{if} its p-value $> \alpha_1$ and its two neighbouring blocks have not been merged at the current step,
merge two neighbouring blocks as one and update the layer label of their inner boundary
by one.
\newline
\indent \textbf{for} each new boundary $t_m$ \newline
\indent \indent recalculate the GLR test and obtain the p-value \newline
rank p-values in descending order.
\medskip
\hrule
\medskip

\subsection*{Validating performance by simulation studies}

Hi-C data have random interactions generated in random ligation of DNA segments \cite{Dixon:2016, Rao:2015}. Polymer models show a decrease in the random interaction score as the distance between two loci increases. Because there is no true answer for TAD boundaries in real Hi-C datasets for validation, we first tested HiCKey on two simulation datasets that were originally created for assessing several computational methods by Forcato \cite{Forcato:2017}. These datasets were simulated by a quasi-negative-binomial generator modified from Lun \cite{Lun:2015}, with each $y_{ij}$ specifically designed to approximate real Hi-C data well. Using the same datasets facilitates the comparison between HiCKey and other methods. The specific datasets we used are as follows:
\begin{itemize}
\item \indent [(Sim1)] {\it Matrices without nested TADs}. It consists of 20 simulated Hi-C matrices with noise levels of 4\%, 8\%, 12\% and 16\%. These matrices contain no nested TAD structure, and each matrix has a size of approximately 4,500 with 171 diagonal blocks.
\item \indent [(Sim2)] {\it Matrices with nested TADs}. It consists of 20 simulated Hi-C matrices with noise levels of 4\%, 8\%, 12\% and 16\%. These matrices differ from Sim1 in that they contain nested TADs. In particular, each matrix has a size of approximately 4,500 and contains 910 diagonal blocks with three layers of hierarchical structure.
\end{itemize}
A previous study \cite{Lun:2015} reported that the noise level, which they refer to as the biological coefficient of variation, varies between 0\% and $16\%$.

The performance was evaluated by four measures. First, the true positive rate (TPR) was defined as the number of detected true boundaries divided by the number of total true boundaries. Second, the false discovery rate (FDR) was defined as the number of falsely detected boundaries divided by the total number of detected boundaries. Third, the difference between the estimated and true number of TAD boundaries was defined as $\widehat{K}-K$. If there were several matrices in a simulation dataset, we calculated the average of all $\widehat{K}-K$ of the matrices. Fourth, to evaluate the consistency between true hierarchical TAD structures and HiCKey TADs, we calculated the Fowlkes-Mallows index ($B_k$) \cite{Fowlkes:1983}, where $k$ is the hierarchical level. Since there are three levels of hierarchical structures embedded in the Sim2 dataset, $B_1$, $B_2$ and $B_3$ were calculated for each hierarchical level from the bottom layer $B_1$ to the outer layer $B_3$. It was noted that the $B_k$ index lies between $0$ and $1$. If two partitions were perfectly matched, then $B_k = 1$. For each noise level, we calculated the average score of $B_k$ for 1,000 random initializations. To obtain the Fowlkes-Mallows indices under the null hypothesis that the two clusterings are unrelated, we calculated control $B_k$ between the true hierarchical structures and randomly relabelled HiCKey TADs (relabelling) by using the Fowlkes and Mallows formula \cite{Fowlkes:1983}.

\subsection*{Validating the robustness of HiCKey by simulation studies}

To evaluate the robustness of HiCKey against different initial boundaries (change-points), we performed validations on simulated datasets Sim1 and Sim2 and real datasets of hESC and IMR90 cell lines. For each dataset, we first ran HiCKey ordinarily and recorded the number of detected change-points as well as their locations. Then, we ran HiCKey 1,000 times with random selection of the first change-point in the whole matrix. We evaluated the result consistency of randomly starting and the ordinary run by using the criteria TPR and $\widehat{K}-K$.

To test the performance of HiCKey on datasets with different distributions, we generated simulation matrices whose entries followed Gaussian and NB distributions. We considered the following two scenarios for the Hi-C matrix of size $500\times 500$:

\begin{itemize}
\item \indent [(Sim3)] {\it Gaussian distribution.}
Let $K=31$ (change-point numbers), and their locations were uniformly drawn with the smallest block size that was larger than 4. We set the mean of each element, $u_{ij}$, as $u_{ij} = \mu_k \sim$ Gamma$(4, 18)$ for $(i, j) \in A_{k-1,k}$, $k=1, \dots, 31$, and $\mu_0=0$ for $(i, j) \in A - \cup_{k=1}^{31} A_{k-1,k}$, where the numbers were estimated by real Hi-C data. The values of $x_{ij}$ were generated by
\begin{equation*}
x_{ij} \sim \left\{ \begin{array}{ll}
N(u_{ij}, \sigma^2) & \mbox{for } (i, j) \in A_{k-1,k}, \\
\max \{ N(0, \sigma^2), 0 \} & \mbox{o.w.}
\end{array} \right.
\end{equation*}
\item \indent [(Sim4)] {\it Poisson and negative binomial distributions.}
Let $K=31$. Change-point locations and element means $\mu_k$ were similarly generated as (Sim3). Furthermore, $x_{ij}$ was generated by an NB model \cite{Lun:2015}
\begin{equation*}
x_{ij} \sim \left\{ \begin{array}{ll}
NB( \nu^{-1}, (1+\nu \mu_k)^{-1}) & \mbox{for } (i, j) \in A_{k-1,k}, \\
0.5 {\bf 1}_{\{0\}} + 0.5 NB( \nu^{-1}, (1+\nu \cdot \min_{k} \{ \mu_k\})^{-1})
& \mbox{o.w.}
\end{array} \right.
\end{equation*}
\noindent Note that $x_{ij}$ in the complementary region $A - \cup_{k=1}^{K+1} A_{k-1,k}$ follows a mixture of point mass and NB distribution.
\end{itemize}

$NB( \nu^{-1}, (1+\nu \mu_k)^{-1})$ provides an NB distribution with mean $\mu_k$ and variance $\mu_k + \nu \mu_k^2$. The parameter $\sqrt{v}$, referred to as the {\it biological coefficient of variation} (BCV) \cite{Lun:2015}, varies from 0 to 16\%. Hence, we set $\sqrt{\nu} = 0, 0.05, 0.10, 0.15$ in (Sim4). Note that $\nu=0$ corresponds to the case in which $x_{ij}$ follows a Poisson distribution with mean $\mu_k$. In (Sim3), $\sigma^2$ was specified as $\sigma^2 \approx 72+72^2 \cdot \nu$, where 72 was the mean of Gamma$(4, 18)$.

\subsection*{Hi-C datasets for real data case studies}

High-resolution \emph{in situ} Hi-C data of seven cell lines produced by Rao \cite{Rao:2015} were downloaded from the Gene Expression Omnibus (GEO) database (http://www.ncbi.nlm.\newline nih.gov/geo/) with the accession number GSE63525. We applied HiCKey on 25 kb resolution Hi-C data for all seven cell lines, which included GM12878, HMECs, HUVECs, IMR90 cells, K562 cells, KBM7 cells, and NHEKs. In addition, we downloaded their predicted TADs using the Arrowhead method \cite{Rao:2015} and denoted them by Rao-TAD in the rest of the paper. We also downloaded early Hi-C data generated by Dixon \cite{Dixon:2012} for two human cell lines, HESC and IMR90, whose resolutions are 40 kb. Histone modifications and TF binding peaks were obtained from ENCODE and the Roadmap epigenomics project using the WashU genome browser \cite{new25}.

\subsection*{Comparing TADs across samples}

Since HiCKey outputs p-values of TAD boundaries, they can be extended to compare boundary differences across cell lines. For a boundary, $m$, we calculated its p-values, $p_1(m)$ and $p_2(m)$, in two different samples of Hi-C matrices. Assuming two Hi-C experiments are independent, we used Fisher's method \cite{fisher} to combine two p-values into one test statistic $\chi_4^2:= -2\ln(p_1(m))-2\ln(p_2(m))$. Here, $\chi_4^2$ follows a chi-squared distribution of 4 degrees of freedom, and its p-value is denoted by $p_f$. The p-value $p_f(m)$ will decrease if $p_1(m)$ and/or $p_2(m)$ decrease.

\subsection*{Memory and running time optimization}

Recent \emph{in situ} Hi-C experiments have generated datasets with a resolution as high as 1 kb, resulting in large matrices. Therefore, it is essential to optimize memory usage and running time. Some Hi-C data processing pipelines, such as HiC-Pro \cite{new26}, place much emphasis on memory efficiency. Many existing TAD detection methods use a matrix as input; however, this is infeasible for high-resolution data. For example, storing a single Hi-C matrix of 5 kb resolution under double precision might require 18 G memory. We used several strategies for computing resource optimization.

First, high-resolution Hi-C matrices are sparse, as most elements are zero. Our program can read both matrix or list forms (Rao's data consist of non-zero elements and their indices \cite{Rao:2015}). HiCKey stores only non-zero elements in the upper triangular part of the Hi-C matrix. Second, the top-down binary segmentation is the most time-consuming step of HiCKey. To calculate all $Z_m$ in ${\bf X}=(x_{ij}) \in \mathbbm{R}^{n\times n}$, we first calculate the sums of every row and column. As $m$ moves from $1$ to $n$, only linear arithmetic operations are needed.

Second, for the best case in which each change-point is allocated in the middle of matrix $A$, we need at most $log_2 n$ iterations with operations $O(n^2)$. In the worst case, each iteration generates one sub-matrix as small as possible and the other as large as possible. This results in at most $n/\xi$ iterations with operations $O(n^3)$ (see detailed calculations in Additional file 1). All the tests were conducted on a regular laptop with an Intel(R) Core(TM) i5-7200U CPU with a 2.50 GHz processor and 12 GB memory.

\section*{Results}

\subsection*{Performance of HiCKey in detecting TADs}

\begin{figure}[h!]
\caption{The performance of four methods.}
\includegraphics[width=0.96\textwidth]{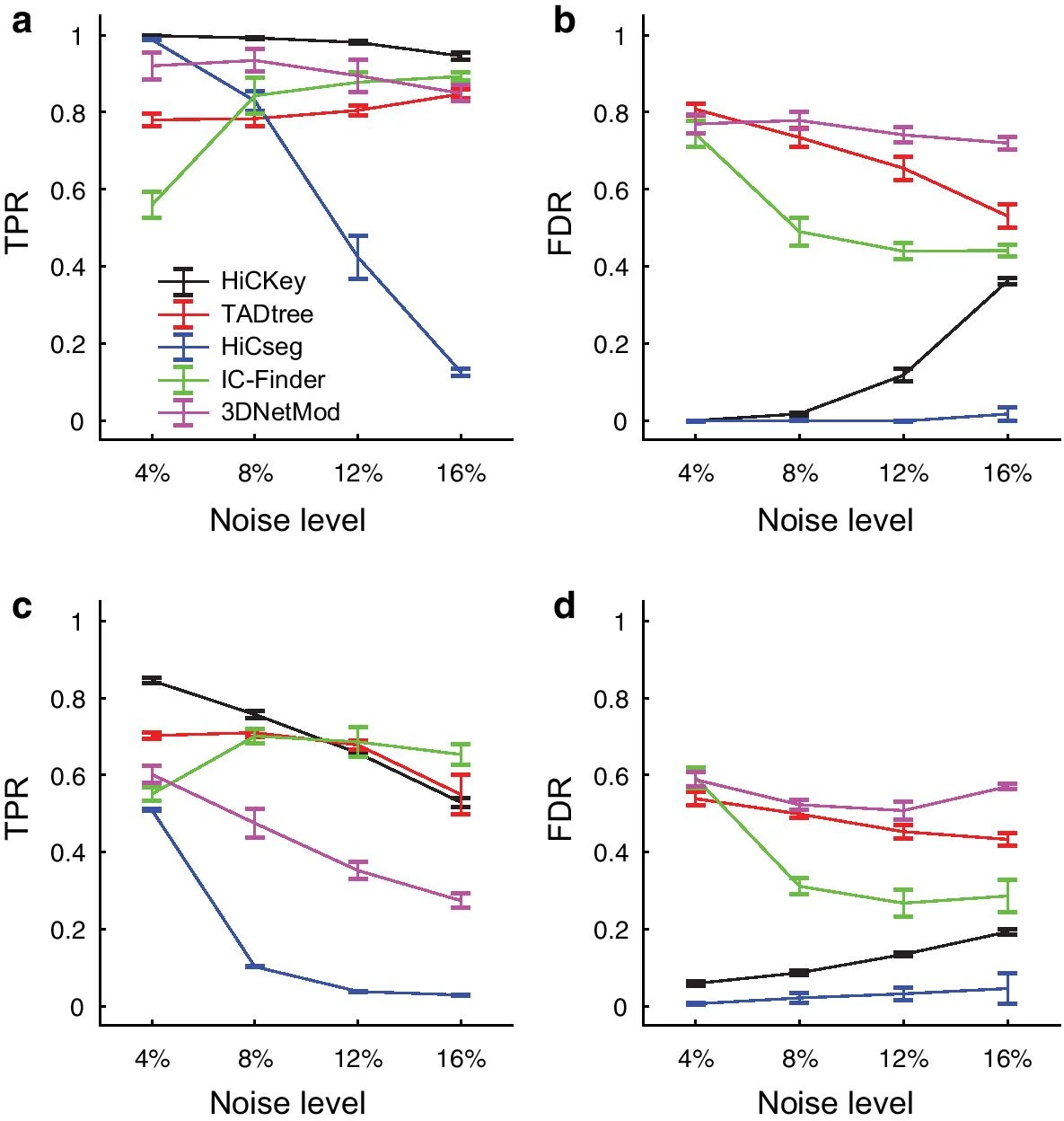}\par
\textbf{a.} TPR of Sim1 dataset. \textbf{b.} FDR of the Sim1 dataset. \textbf{c.} TPR of the Sim2 dataset. \textbf{d.} FDR of the
Sim2 dataset.
\label{Fig3_Performance}
\end{figure}

To understand the performance of HiCKey in detecting TADs, we first tested it on large-scale simulated Hi-C matrices (Forcato, Nicoletti et al. 2017), which included two types of data, Sim1 without nested TAD structures and Sim2 with nested TAD structures. For both Sim1 and Sim2, four noise levels of 4\%, 8\%, 12\% and 16\% were added to test the robustness of HiCKey against the random collision noise of interactions. Testing at different noise levels is critical since large numbers of random collision interactions are observed in HiC data (Rao, Huntley et al. 2014, Dixon, Gorkin et al. 2016). First, on dataset Sim1, HiCKey achieved a high TPR of 0.9988 under a 4\% noise level (Fig. \ref{Fig3_Performance}. Additional file 1, Table S1). As the noise level increased from 4\% to 16\% (4-fold change), the TPR decreased to 0.9459 (0.947-fold change). The fold change ratio of TPR and noise was 0.24 (0.947/4), indicating that the TPR of HiCKey was robust against noise changes. When the noise level increased from 4\% to 8\% and 12\%, the FDR slightly increased from 0 to 0.0173 and 0.1176, respectively. Additionally, the FDR increased to 0.3618 at the 16\% noise level. We also compared the number of TADs estimated by HiCKey with the true value. We found that HiCKey produced a very accurate average number of TADs at the 4\% noise level ($\widehat{K}-K=-0.2$). As the noise level increased, the estimated number of TADs increased ($\widehat{K}-K$ as 1.8, 19.4 and 82.6 for noise levels 8\%, 12\% and 16\%, respectively). We then tested HiCKey on dataset Sim2. At a 4\% noise level, HiCKey achieved 0.845 TPR and 0.059 FDR (Additional file 1, Table S2). When the noise level increased to 8\% and 12\%, the TPR decreased to 0.757 and 0.6568, respectively. We noticed that TPR and noise level were linearly correlated ($R^{2}=0.9926$), demonstrating that HiCKey can remain stable with these noise changes. In summary, these validation results at different noise levels suggest that HiCKey is robust against random collision noise, especially for noise levels ranging from 0 to 12\%.

We calculated the Fowlkes-Mallows index $B_k (k=1,2,3)$ \cite{Fowlkes:1983} to evaluate the consistency between true hierarchical TAD structures and HiCKey TADs, as there are three layers of hierarchical structures embedded in the Sim2 dataset. The $B_k$ index lies between $0$ and $1$, where larger index scores indicate higher similarities among two compared hierarchical structures. Overall, we found that the $B_k$ indices were larger than 0.8196 for all three hierarchical layers at the 4\% noise level (Additional file 1, Table S3). When noise levels increased to 16\%, they decreased slightly but maintained fair scores all larger than 0.6887. At four noise levels, the control Fowlkes-Mallows indices were no more than 0.01 after recalculating $B_k$ between the true hierarchical structures and randomly relabelled HiCKey TADs (Additional file 1, Table S3). We also found that the $B_2$ and $B_3$ indices were more stable than $B_1$ when noise levels increased, suggesting that HiCKey is more robust against noise in detecting second and third levels of TAD structures.

We examined practical running time and memory usage on a real Hi-C matrix of chr1 (the largest among 23 chromosomes) in the GM12878 cell line \cite{Rao:2015}. Under resolutions of 50 kb, 25 kb, 10 kb and 5 kb, the running times of HiCKey were 23 s, 53 s, 106 s and 157 s, respectively. Additionally, the practical running time was approximately $O(n^2)$ (Additional file 1, Fig. S2a). Its memory usage was 170 Mb, 389 Mb, 768 Mb and 980 Mb, respectively, and was also approximately $O(n^2)$ (Additional file 1, Fig. S2b). Taken together, HiCKey requires reasonable computing resources in processing high-resolution Hi-C matrices.

\subsection*{Robustness against different distributions and initialization}

In Hi-C data analysis, the read counts of interactions were usually assumed to follow an NB distribution \cite{Rao:2015, Carty:2017, Kate:2020} or Poisson distribution \cite{Mohamed:2018} or normalized data \cite{new19}. However, these distribution models cannot fully capture the characteristics of chromatin interactions in Hi-C experiments due to the divergent confounding factors observed in real biological systems, which results in a complicated mixed model \cite{Dekker:2002, Forcato:2017, nobel}. In HiCKey, we derived a GLR test that can be broadly used for multiple distributions but is not limited to the NB distribution. To test the performance of HiCKey on different distributions, we simulated 1,000 interaction matrices with normal and NB distributions (see details in the Methods section). At four different noise levels, we found that the TPRs were all larger than 0.99, while the FDRs were less than 0.0055 (Additional file1, Table S4), suggesting that HiCKey is robust against different distributions.

To test whether HiCKey is sensitive to the initial choice of change-point allocation, we constructed new validations by randomly selecting the initial location. We performed validations on simulation datasets Sim1 (Additional file 1, Table S5) and Sim2 (Additional file 1, Table S6) and real datasets of hESCs (Additional file 1, Table S7) and IMR90 cell lines (Additional file 1, Table S8). Regarding these validations, TPR rates were all larger than 90\%, while the means of $\hat{K}-K$ were very small, indicating that HiCKey is robust against the initial boundary selection.

\subsection*{Comparisons with other methods}

We compared HiCKey with four popular methods, HiCSeg \cite{Levy-leduc:2014}, TADtree \cite{Weinerb:2015}, IC-Finder \cite{Haddad:2017} and 3DNetMod \cite{3dnetmod}, on simulation datasets Sim1 and Sim2. For Sim1, HiCKey achieved not only higher TPRs at four noise levels but also slow declines (Fig. \ref{Fig3_Performance}a and Additional file 1, Table S1). The FDR of HiCKey at 16\% noise was smaller than those of IC-Finder, TADtree and 3DNetMod (Fig. \ref{Fig3_Performance}b). We found that HiCSeg tended to retrieve large TADs, resulting in fewer detected TAD boundaries. For example, HiC-Seg’s $\widehat{K}-K$ were -2.2, -29.2, -99.2 and -150.2 at noise levels of 4\%, 8\%, 12\% and 16\%, respectively, explaining why their FDRs were always lower but TPRs decreased sharply. TADtree and IC-Finder outputted more TADs than the true value at the 4\% noise level ($\widehat{K}-K$ = 397.6 and 207.4, respectively), but the number of falsely identified TADs decreased with increasing noise levels. 3DNetMod tended to output more TADs than the true numbers at four noise levels and thus had higher FDR rates. For Sim2, HiCKey achieved the highest TPRs at noise levels of 4\% and 8\% but slightly dropped below TADtree and IC-Finder at noise levels of 12\% and 16\% (Fig. \ref{Fig3_Performance}c and Additional file 1, Table S2). However, TADtree, 3DNetMod and IC-Finder suffered from much higher FDRs (Fig. \ref{Fig3_Performance}d). Taken together, HiCKey achieved good performance, especially for lower noise levels.

\begin{figure}[h!]
\caption{True and estimated hierarchical structures of two samples in the Sim2 dataset.}
\includegraphics[width=0.96\textwidth]{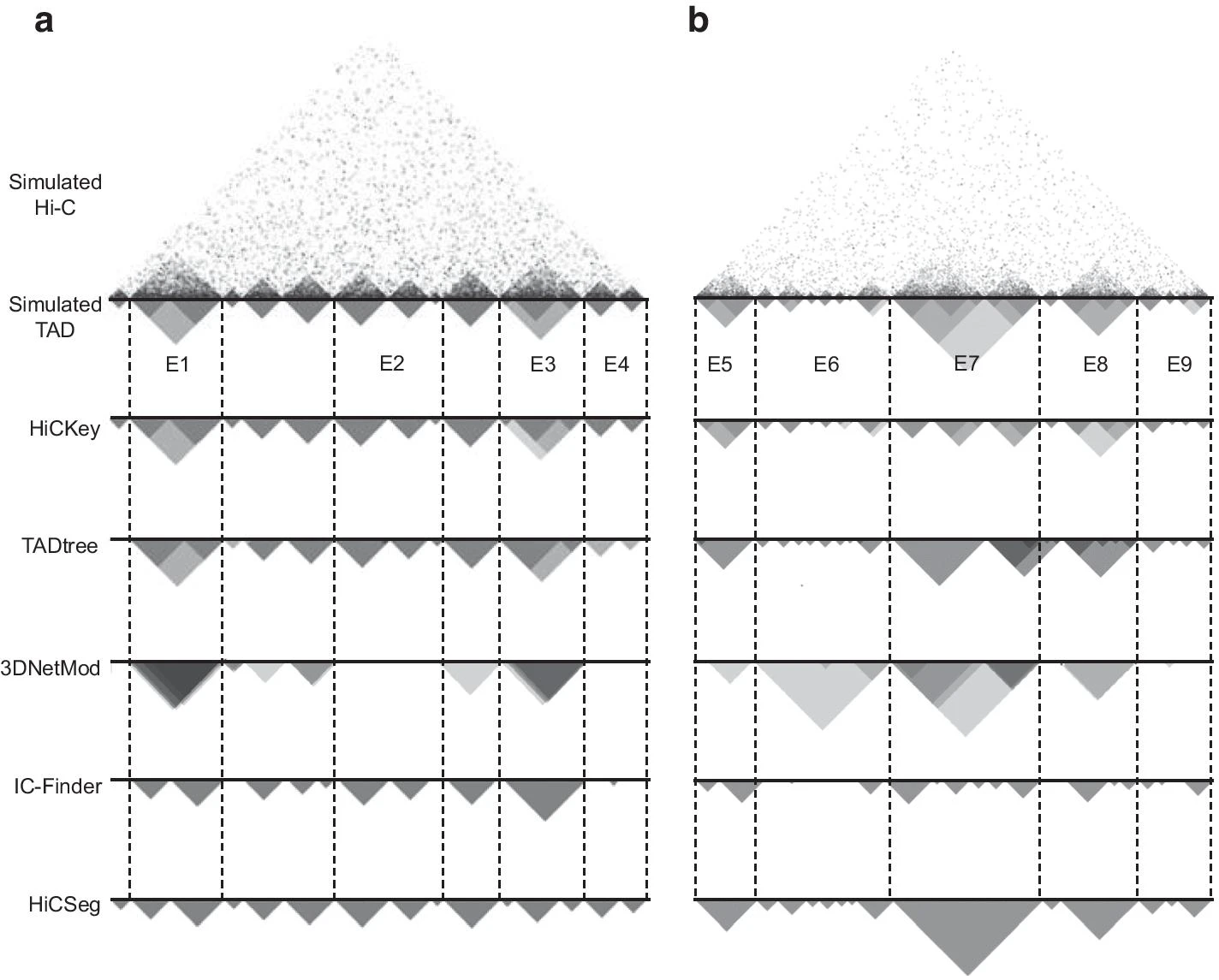}\par
\textbf{a.} Simulated case1, chr: 75600000-82720000. \textbf{b.} Simulated case2, chr: 127000000
-137000000.
\label{Fig4_SimulationCase}
\end{figure}

We specifically investigated two regions with hierarchical TADs in dataset Sim2 that were used for comparative analysis in a previous study \cite{Forcato:2017} (Fig. \ref{Fig4_SimulationCase}). The results showed that HiCKey, TADtree and 3DNetMod can detect single and hierarchical TADs for both cases, while IC-Finder and HiCSeg can only output bottom single TADs. At E1-E6, we found that HiCKey can correctly detect not only all the TAD boundaries but also their hierarchical organization, while TADtree and 3DNetMod made a few false predictions (E1, E3, E6) or missed outputting bottom TADs (E2, E4, E9). Region E7, consisting of complicated hierarchical structures, was a challenge for HiCKey and TADtree. Although HiCKey correctly detected all the boundaries in E7, it missed the outer layer. TADtree missed several boundaries and wrongly merged a neighbouring TAD into the block. 3DNetMod showed good hierarchical details at E7 and reported a large hierarchical TAD at E6, which seems to have no clear interaction blocks. Overall, these genome-wide analyses and detailed examples showed that HiCKey has good performance in detecting TAD boundaries and their hierarchical organization.

\subsection*{Hierarchical architecture of chromosomal organization}

We applied HiCKey to \emph{in situ} Hi-C data of seven cell lines. HiCKey successfully outputted the boundary positions, p-values and hierarchical levels. Following Weinreb and Raphael’s method \cite{Weinerb:2015}, we defined the root TAD as order 1. If a root TAD had sub-level TADs, the two sub-TADs were of order 2. Similarly, sub-sub-TADs were of order 3, and so on. This is the same as how we defined the hierarchical order of TADs in the Methods section. In total, we detected 8,586, 8,200, 8,903, 9,043, 8,801, 6,246 and 7,726 TADs in the GM12878, hMEC, HUVEC, IMR90, K562, KBM7 and NHEK cell lines, respectively (Additional file 2). In each cell line, we compared our results with Rao-TAD. First, we found that their allocations of TADs in the 23 chromosomes were similar ($p$-values $ > 0.23$ in all seven cells, two sides chi-square test). Second, we considered that a Rao-TAD boundary was matched if there was a HiCKey boundary located within its 2-bin (50 kb) distance. The proportions of matches between Rao-TADs and HiCKey TADs were 48.90\%, 65.56\%, 58.71\%, 57.98\%, 52.22\%, 51.14\% and 57.90\% for GM12878, hMECs, HUVECs, IMR90, K562, KBM7 and NHEKs, respectively. Furthermore, we used the hypergeometric test (all the chromosomal bins were taken as population, boundaries of Rao-TAD as successes, HiCKey boundaries as samples, and the matched ones as sampled successes) to measure how significantly HiCKey TAD boundaries matched with Rao-TAD boundaries. On seven cell lines, the p-values were all calculated as less than 1.0e-10, indicating that they were significantly matched.

\begin{figure}[h!]
\caption{Integrative analysis of TAD and histone modifications.}
\includegraphics[width=0.96\textwidth]{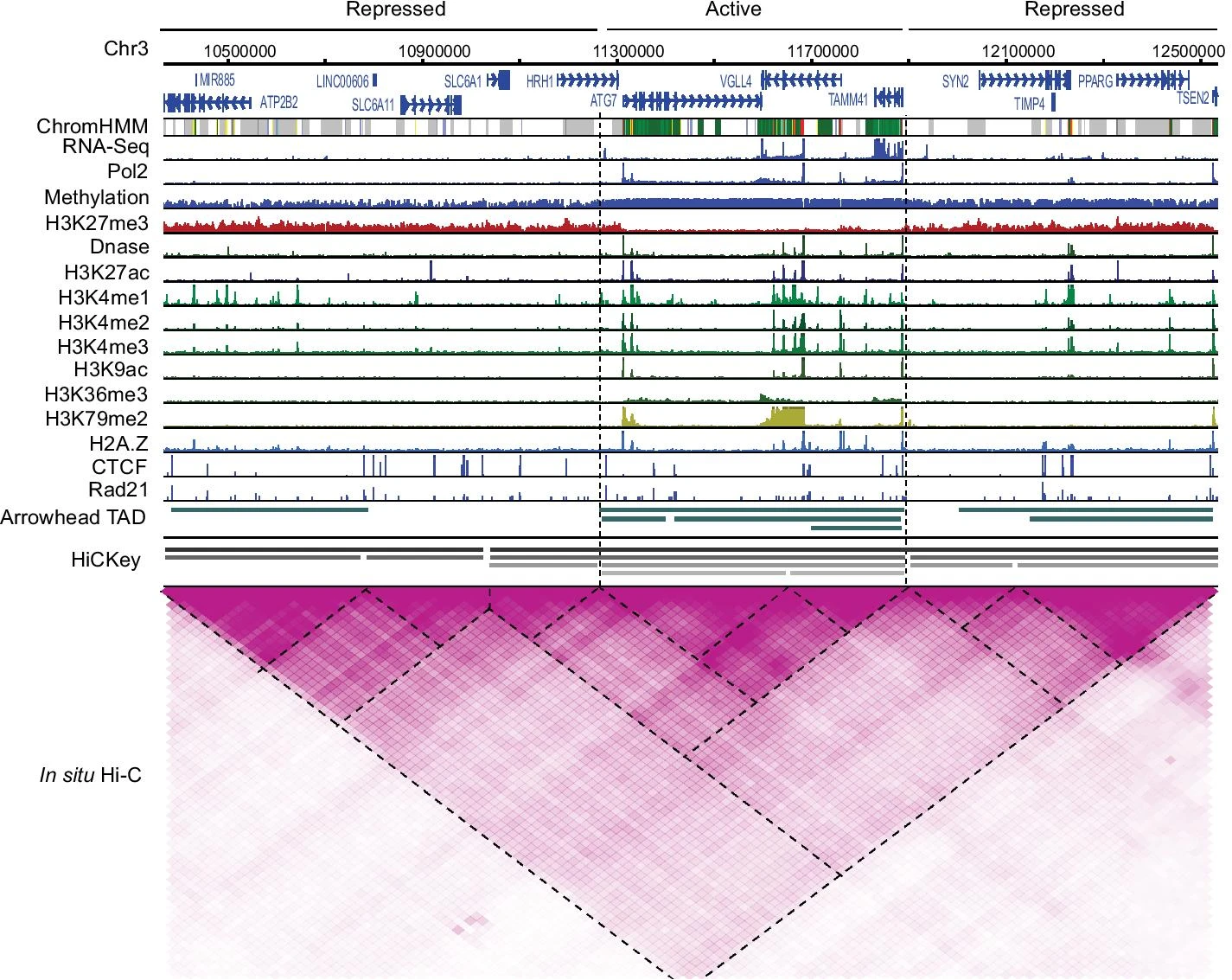}\par
A 2 Mb chr3 region of the GM12878 cell line shows one active region surrounded by two repressive regions. All signals are normalized by the 90-quantile value within this region. The tracks were generated by the WashU genome browser http://epigenomegateway.wustl.edu/legacy/.
\label{Fig5_RealCase1}
\end{figure}

Multiple levels of TADs were detected in each cell line. For example, Fig. \ref{Fig5_RealCase1} demonstrates our estimation of hierarchical TADs and Rao-TAD in a local region of chr3: 10700000-11300000 of GM12878. Overall, HiCKey TAD estimations exhibited more hierarchical layers than Rao-TADs. There was no Rao-TAD within a large sub-region (chromosome 3:10700000-11300000); however, clear blocks were observed from the Hi-C interaction heatmap. Extending the analysis to the genome-wide level, we found that although the highest order of TADs can reach 7, most of them exhibited an order 1 or 2 ($97.86\% \pm 0.71\%$, Additional file 1, Table S9), suggesting that hierarchical TADs are enriched in certain chromosomal regions.

\subsection*{Hierarchical organizations are enriched in active regions compared with repressive regions}

To test whether biased hierarchical organizations at different chromosomal regions are related to different biological insights, we performed an integrative analysis of TAD structures and epigenetic markers. Previous association analysis revealed that neighbouring TADs usually have different histone modification patterns \cite{Chen:2016} and that TAD boundaries are primarily associated with CTCF and Rad21 binding peaks \cite{Dixon:2012, Rao:2015}. Here, we downloaded several histone signals and protein binding peaks for GM12878. We examined a ~$\sim$2 Mb region (chr3: 10700000-11300000, Fig. \ref{Fig5_RealCase1}) that was partitioned into three parts, an active region flanked by two repressed regions. The active region was exhibited by biological signals, such as high RNA-seq signals of genes, Pol2 binding peaks, and multiple histone modifications (e.g., H3K4me3 and H3K27ac). We observed that the active region contained more hierarchical TADs than the repressed regions (Fig. \ref{Fig5_RealCase1}, HiCKey track). 

We next examined genome-wide TAD boundary enrichment in active chromosomal regions. To search genome-wide, we used the active/repressive annotations of chromosomal regions for six cell lines \cite{Chen:2016} and compared the numbers and layers of boundaries between them. First, we confirmed that among all six cell lines, TAD boundaries were enriched in active regions compared with repressive regions ($p$-value $< 0.01$, one-sided Fisher exact test, Additional file 1, Table S10). For instance, among the estimated 8,200 active/repressive TAD boundaries in the IMR90 cell line, 5,494 were located in active regions (64,220 bins of 40 kb), while only 2,706 were in repressive regions (46,400 bins of 40 kb). Next, we checked the layer annotation of boundaries. By comparing the layer number distributions, we found that they were significantly different, as the average layer number in active regions was larger than that in repressive regions ($p$-value $< 0.01$, K-S test). Thus, these genome-wide analysis results demonstrated that active chromosomal regions usually contain more TAD boundaries and richer TAD structures, indicating that active regions may employ more precise spatial organizations to regulate gene expression.

\subsection*{Detecting conserved and dynamic TAD boundaries between cells}

Conserved and dynamic boundaries, as a result of cell-specific gene expression organization/regulation, different developmental conditions, or chromosomal variants in diseases, can be compared by Hi-C data of two different samples \cite{Dixon:2016, Spielmann:2018}. Here, we examined the 8,801 TAD boundaries of the K562 myelogenous leukaemia cell line and 9,043 TAD boundaries of the IMR90 normal human fibroblast cell line detected by HiCKey. First, we found that 7,286 boundaries were co-localized within a 2-bin distance. These high proportions of matched boundaries in K562 (82.79\%) and IMR90 (80.57\%) cells are consistent with early observations that TADs are conserved among mammalian cells \cite{Dixon:2012, Dixon:2016, Rao:2015}. Second, among the co-localized boundaries, 7,280 of them have Fisher's combined p-values $p_f < 0.01$, indicating that they are conserved in both cell lines. We also detected 1,621 K562 TAD boundaries changed in IMR90, and 1,763 IMR90 TAD boundaries changed in the K562 cell line, providing potential candidates for TAD boundaries that may be involved in cell-specific regulation.

\begin{figure}[h!]
\caption{\bf Different TAD structures of chr2:38900000-43100000 in K562 and IMR90.}
\includegraphics[width=0.96\textwidth]{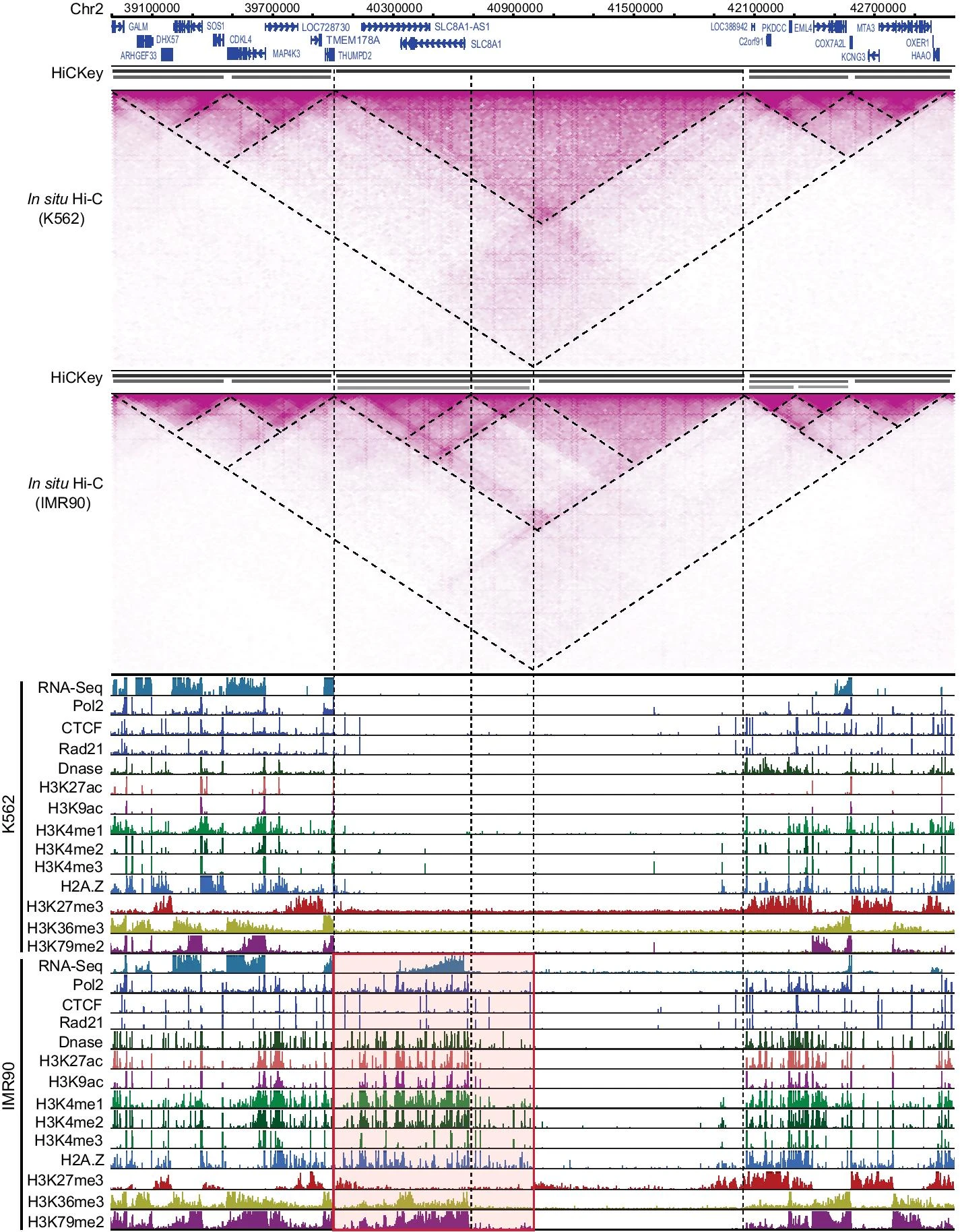}\par
A chr2:40000000-41000000 region showed hierarchical TADs in IMR90 but not in K562. ChIP-seq signals are normalized by the 90-quantile value within this region. The tracks were generated by the WashU genome browser http://epigenomegateway.wustl.edu/legacy/.
\label{Fig6_RealCase2}
\end{figure}

To demonstrate dynamic changes in TADs and their potential biological functions related to gene transcription, we investigated a large chromosomal region (~$\sim$4.2 Mb, chr2:38900000-43100000) of the K562 and IMR90 cell lines as representative. In this region, three large TADs were estimated in both K562 and IMR90, but a TAD (chr2:40000000-42050000) showed novel hierarchical sub-TAD structures in IMR90 but not in K562 (Fig \ref{Fig6_RealCase2}). Moreover, two sub-level TADs (chr2:40000000-41000000 and chr2:41000000-42050000) were observed in IMR90 with clear Hi-C interaction patterns. In contrast, their interactions were uniform in K562 cells. Furthermore, two smaller blocks (chr2:40000000-40700000 and chr2:40700000-41000000) were detected in the sub-TAD (chr2:40000000-41000000) of IMR90. The smaller block (chr2:40000000-40700000) included the protein-coding gene SLC8A1 and the lncRNA gene SLC8A1-AS1. SLC8A1 (solute carrier family 8 member A1) is a protein-coding gene linked to multiple diseases, such as long Qt syndrome 9, cardiac diseases and aromatase deficiency \cite{new1, new2, new3}. We found that this small block was active in IMR90 but not in K562 by several signals. First, SLC8A1 was highly expressed in IMR90 cells but not in K562 cells. This was consistent with the novel binding signals of Pol2, H2A. Z, CTCF and Rd21 in IMR90. In addition, novel histone modifications were observed in IMR90 but not in K562. Another small block (chr2:40700000-41000000) was at the 5’ region of the SLC8A1 gene, but no genes were included. It also contained novel signals of CTCF and Rad21, suggesting that its generation may be mediated by CTCF and Rad21 to regulate SLC8A1 expression. Overall, these results demonstrate that HiCKey can detect not only TAD hierarchies but also their difference across samples. The hierarchical TAD structures in IMR90 further support our findings that active regions of chromosomes usually contain more and richer TAD structures for regulating gene expression.

\section*{Discussion}

The identification of TADs and their hierarchical structures is extremely important in the study of chromatin interactions. We developed a novel GLR test to detect change-points in Hi-C matrices and studied its asymptotic properties. Based on the GLR test, we introduced HiCKey to decipher the hierarchical structure of TADs. The performance of HiCKey is endorsed by extensive simulation and real data analysis. The retrieved hierarchical TADs are consistent with diverse biological signals, including histone modification and ChIP-seq data. We further found much more detailed TAD structures in active chromosomal regions. Comparative analysis of TADs across various cell lines revealed that different TAD organizations harbour disease-related genes, providing insights into how disordered interactions are linked to different cancer types.

\subsection*{Considerations for different distributions of chromatin interactions}

To explain chromatin interaction strength decay, polymer models propose that the average pairwise contact probability decreases and asymptotically follows a power law with a given contour distance \cite{new4, new5, new6}. NB is another popular model used not only in the simulation of Hi-C data but also in other types of interaction data, such as ChIA-PET, CAPTURE-seq and HiChIP. Although such methods can approximately explain the interacting decay along the off-diagonal of the Hi-C matrix, they are challenged by several confounding factors in real biological systems. First, most promoter-enhancer interactions are usually regulated by divergent TF proteins or structural proteins, such as Pol2, GATA1/2, CTCF and Rad21. These deterministic factors have been evolutionarily fixed in different cell lines, thereby reducing the randomness of chromatin interactions that is captured in Hi-C experiments. In fact, most promoter-enhancer interactions are observed over short distances (tens of kb) from gene promoters to their neighbouring enhancers. Second, different experimental factors, such as the cross-linking of chromatin, chromatin digestion, and streptavidin pull-down of biotinylated ligations, may also affect the signal/noise ratio of Hi-C data \cite{new7}. Existing distribution models, for example, the NB distribution, cannot fully capture the characteristics of chromatin interactions in Hi-C experiments \cite{new5, new8}. In HiCKey, we derived a GLR test that can be broadly used for multiple distributions and is not limited to NB. This could potentially be very useful for complex Hi-C or ChIP-PET or HiChIP experiments in which interaction strengths may follow some unknown distribution.

\subsection*{Phase transition, loop extrusion and chromosomal hierarchies}

Although Hi-C data provide a landscape of interacting strengths among chromosomes, mechanistic explanations of how chromatin interactions are dynamically formatted and regulated are lacking. At large scales (e.g., A/B compartments and TADs), it seems that interactions can be self-organized by levels of epigenetic modifications that are formalized as the phase transition in the neighbourhood of typical physiological conditions \cite{new6, new9, new10}. To further consolidate this phase-transition model, it will be highly desirable to manipulate \textit{in vivo} histone-tail modifications for comparative Hi-C analysis. At smaller scales and active chromosomal regions, the loop extrusion model is proposed to link the CTCF and cohesin proteins into the formation of local looping structures \cite{new11, new12, new13}. The loop extrusion model suggests that structural maintenance of chromosome proteins (cohesin or condensin) progressively extrudes chromatin until it is blocked by CTCF bound to properly oriented site pairs \cite{new14, new15, new16}. In our comparative studies of K562 and IMR90 cell lines, we found that CTCF and Rad21 binding sites were remarkably changed within the newly established TADs (Fig. \ref{Fig6_RealCase2}), indicating that CTCF-loop domains and enhancer-promoter interactions may be established via an extrusion process involving cohesin and CTCF. In summary, these observations and studies suggest that the phase-transition model and loop extrusion model take place at different scales and chromatin states.

\subsection*{Integrating other Omics data}

With the hierarchical organization of chromatin interactions available, we can deeply investigate the biological functions or principles of how histone modifications are coordinately used, as well as how gene expression is dynamically regulated. Since histone modifications, TF binding sites, and gene expression have been collected for hundreds of cell lines in ENCODE and Roadmap Epigenomics projects, we integrated them for locus-specific and genome-wide analysis. As shown in Fig. \ref{Fig5_RealCase1}, we demonstrated that TADs estimated by HiCKey are consistent with histone modifications, including H3K27ac, H3Kme1/2/3 and many others. Gene expression from RNA-seq and TF binding peaks (Pol2, CTCF and H2A. Z) also confirmed the active and repressive compartment of chromosomal regions. By using multiple omics signals, we further observed different signal patterns in a comparative analysis of K562 and IMR90 (Fig. \ref{Fig6_RealCase2}). We found that the newly expressed genes at active regions in IMR90 were occupied by multiple active histone signals. The new binding peaks of CTCF and Rad21 suggest that these structural proteins may be involved in the local chromosomal conformation for SLC8A1 gene expression in IMR90. Both examples indicate that integrative analysis of multiple omics data and hierarchical organizations is a promising method to fully understand chromosomal compartments and functions. \newline

\section*{Conclusions}

In this work, we presented an efficient method, HiCKey, for detecting and comparing hierarchical TAD structures in Hi-C datasets. We especially derived a GLR test that worked for general distributions. The theoretical results of the GLR test can be used in similar experimental data (such as HiChIP, ChIA-PET and Drop-seq), whose signal may not fully follow the NB distribution but more general mixture distributions. HiCKey was evaluated by using large simulation data and real Hi-C data of mammalian cell lines. First, large-scale validations on simulation data (with or without nested Hi-C structures) show that HiCKey has good precision in recalling known TADs and is robust against random collision noise of chromatin interactions. Second, HiCKey was successfully applied to in situ Hi-C data of seven human cell lines, and its predictions are supported by diverse epigenetic markers and exhibit novel biological discoveries. We concordantly identified multiple layers of TAD organization among these cell lines. In particular, TAD boundaries were found to be significantly enriched in active chromosomal regions compared to repressed regions. HiCKey was manipulated by C++ language for high operation speed. It accepts multiple input formats of the Hi-C matrix and is optimized for processing large matrices constructed from high-resolution Hi-C experiments. With more Hi-C and similar experimental datasets available, we believe our method and theoretical framework will highly inspire computational biologists to design novel pipelines by using the GLR test to elucidate the hierarchical organization of locus-specific chromatin interactions in mammalian genomes or other types of deep sequencing data analysis.

\section*{Abbreviations}
\textbf{TAD:} Topologically Associating Domain\newline
\textbf{GLR:} Generalized Likelihood Ratio\newline
\textbf{NB:} Negative Binomial\newline
\textbf{MCMC:} Markov Chain Monte Carlo\newline
\textbf{TPR:} True Positive Rate\newline
\textbf{FDR:} False Discovery Rate\newline
\textbf{GEO:} Gene Expression Omnibus\newline
\textbf{ENCODE:} Encyclopaedia of DNA Elements

\section*{-Declarations-}
\section*{Ethics approval and consent to participate}
Not applicable

\section*{Consent for publication}
Not applicable

\section*{Availability of data and material}
HiCKey is coded in C++. HiCKey is open source available in the GitHub repository (https://github.com/YingruWuGit/HiCKey).

\section*{Competing interests}
The authors declare that they have no competing interests.

\section*{Funding}
Publication costs are funded by Rowan University Startup grant, 2019, (PI, Yong Chen). Moreover, research reported in this project was partially supported by the NIH grant R01MH109616, the Cecil H. and Ida Green Endowment, the SKR$\&$DPC grant (2017YFA0505503) (PI, Michael Q. Zhang), and NSF DMS-1612501 (PI, Haipeng Xing). The funders had no roles in the design of the study and collection, analysis, and interpretation of data or in writing the manuscript.

\section*{Authors' contributions}
HX, YC and MQZ initiated the concept and supervised the study. HX, YW and YC designed the methodology. YW and YC performed the data analysis. YW implemented the software. YC, YW and HX drafted and reviewed the paper. All authors have read and approved the final manuscript.

\section*{Acknowledgements}
We would like to thank Catherine Tang and Nakoa Kristen Webber for critical reading. \newline
We would like to thank Dr. Mattia Forcato for providing detailed information helping with the comparison of HiCKey with other methods.\newline
We thank the four anonymous reviewers for their insightful suggestions. \newline

\section*{Proof of Theorem 1 and 2}

\noindent {\bf{Theorem 1}}.

Before we prove Theorem 1, let's take a look at Gaussian distribution 
assumption. Instead of negative binomial we assume:

$$
x_{ij} \sim N(\mu_k, \sigma^2), \quad 1\le i \le j \le n, \quad
\mu_k = \left\{ \begin{array}{ll}
\mu_1, & \mbox{if} (i,j) \in A_{0,1}\\
\mu_2, & \mbox{if} (i,j) \in A_{1,2}\\
\mu_0, & \mbox{if} (i,j) \in R_{0,1,2}
\end{array} \right.
$$

\noindent Without loss of generality we take $\sigma$ constant in 
the local region $A$, as they can be scaled to be equal. So we have 
log GLR test statistics:

$$
GLR_{G,m} = \frac{1}{2\sigma^2} \Big(
\frac{S_{A_1}^2}{|A_{0,1}|} + \frac{S_{A_2}^2}{|A_{1,2}|}
+ \frac{S_R^2}{|R|} - \frac{S_A^2}{|A|} \Big)
$$

\noindent some algebra shows directly that $GLR_{G,m} = Z_m$. \newline

Now we come back to negative binomial case. 

\begin{eqnarray}
&& GLR_{NB,m} = \sum_{k=1,2} \Big\{ S_{A_k} \log \Big(
\frac{S_{A_k}/|A_{k-1,k}|}{r+S_{A_k}/|A_{k-1,k}|} \Big)
 \nonumber \\
&& \hspace{50pt} + r |A_{k-1,k}| \log \Big( \frac{r}{r+S_{A_k}/|A_{k-1,k}|} \Big) \Big\}
 \nonumber \\
&& \hspace{50pt} + \Big( S_R \log \Big( \frac{S_R/|R|}{r+S_R/|R|} \Big)
+ r |R| \log \Big( \frac{r}{r+S_R/|R|} \Big) \Big)
 \nonumber \\
&& \hspace{50pt}
- \Big( S_{A} \log \Big( \frac{S_{A}/|A|}{r+S_{A}/|A|} \Big)
+ r |A| \log \Big( \frac{r}{r+S_{A}/|A|} \Big) \Big) \label{glrnb}
\end{eqnarray}

\noindent Under null hypothesis, consider $x_{ij}$ from region $A$ 
in \ref{glrnb}. By central limit theorem we have:

\begin{equation}
S_A/ |A| = \frac{\phi r}{1-\phi} + O_p(|A|^{-1/2}) \label{clt}
\end{equation}

\noindent where $E(S_A/ |A|) = \frac{\phi r}{1-\phi}$. By Taylor expansion around the mean:

\begin{eqnarray*}
&&(S_A/|A|)\log \Big(\frac{S_A/|A|}{r+S_A/|A|}\Big) = 
\frac{\phi r}{1-\phi} \log (\frac{\phi}{1-\phi}) \\
&& \hspace{120pt} + (1-\phi+\log\phi)\Big(
S_A/|A|-\frac{\phi r}{1-\phi}\Big) \\
&& \hspace{120pt} + \frac{1}{2}\Big( \frac{(1-\phi)^2}{\phi r} - \frac{(1-\phi)^2}{r} \Big) 
\Big( S_A/|A|-\frac{\phi r}{1-\phi} \Big)^2 \\
&& \hspace{120pt} + o_p(|A|^{-1})
\end{eqnarray*}
\begin{eqnarray*}
&&\log \Big( \frac{r}{r + S_{A}/|A|} \Big) = \log (1-\phi) - \frac{1-\phi}{r} 
\Big( S_{A}/|A| - \frac{\phi r}{1-\phi} \Big) \\
&& \hspace{80pt} + \frac{(1-\phi)^2}{2r^2} \Big( S_{A}/|A| - \frac{\phi r}{1-\phi} \Big)^2 + o_p( |A|^{-1}).
\end{eqnarray*}

\noindent So we have the last term in \ref{glrnb} as:

\begin{eqnarray*}
&& S_A \log \Big( \frac{S_{A}/|A|}{r+S_{A}/|A|} \Big) + r |A| \log \Big( \frac{r}{r+S_{A}/|A|} \Big) \\
&& \hspace{30pt} = |A|(S_A/|A|)\log \Big(\frac{S_A/|A|}{r+S_A/|A|}\Big) + r |A| \log \Big( \frac{r}{r+S_{A}/|A|} \Big) \\
&& \hspace{30pt} = \frac{1}{2} \frac{(1-\phi)^2}{\phi r} \frac{S_A^2}{|A|} + c_1S_A + c_2|A| + o_p(1)
\end{eqnarray*}

\noindent Where $c_1$ and $c_2$ are some constants. Similarly we 
do the same to other terms in \ref{glrnb}. By the fact that $S_A = 
S_{A_1} + S_{A_2} + S_R$ and $|A| = |A_{0,1}| + |A_{1,2}| + 
|R|$, all the first order terms of $S_A$, $S_{A_1}$, $S_{A_2}$, $S_R$, 
$|A|$, $|A_{0,1}|$, $|A_{1,2}|$ and $|R|$ are cancelled. \newline

Because $m/n$ holds constant, $\frac{|A_{0,1}|}{|A|}$, $\frac{|A_{1,2}|}{|A|}$ and $\frac{|R|}{|A|}$ also hold constant. 
We have under the null hypothesis:

\begin{equation}
\frac{2\phi r}{(1-\phi)^2}GLR_{NB, m} =
\frac{S_{A_1}^2}{|A_{0,1}|} + \frac{S_{A_2}^2}{|A_{1,2}|}
+ \frac{S_R^2}{|R|} - \frac{S_A^2}{|A|}+ o_p(1)
\end{equation}
\begin{equation}
GLR_{NB, m} = Z_m + o_p(1)
\end{equation}

\noindent Notice that $\frac{\phi r}{(1-\phi)^2}$ is the variance of 
negative binomial distribution. So under null hypothesis, if we have all the 
elements scaled by the common $\sigma_0$ they have variance equals to 
$1$. \newline

Last, if we assume read counts are Poisson random variables with 
blockwise constant parameter $\lambda$, the GLR test statistics is:

$$
GLR_{P,m} = S_{A_1} \log \frac{S_{A_1}}{|A_{0,1}|} +
S_{A_2} \log \frac{S_{A_2}}{|A_{1,2}|} +
S_R \log \frac{S_R}{|R|}
- S_{A} \log \frac{S_{A}}{|A|}
$$

\noindent By exactly similar arguments we have under null hypothesis: 

\begin{equation}
2\lambda GLR_{P,m} = 
\frac{S_{A_1}^2}{|A_{0,1}|} + \frac{S_{A_2}^2}{|A_{1,2}|}
+ \frac{S_R^2}{|R|} - \frac{S_A^2}{|A|}+ o_p(1)
\end{equation}
\begin{equation}
GLR_{P,m} = Z_m + o_p(1)
\end{equation}

Therefore, The GLR statistics $GLR_{G,m}, GLR_{P,m}, GLR_{NB,m}$ are
asymptotically equivalent to $Z_m$.
\hfill \(\Box\)

\bigskip

\noindent {\bf{Theorem 2}}.

Consider a Gaussian random field $G(s, t)$ defined on the upper triangular part of a unit square, $B = \{ (s, t) | 0 \le s \le t \le 1 \}$. The random field $G(s, t)$ satisfies the following properties: (1) for $s, t \in (0, 1)$ and $s < t$, $\partial^2 G(s, t) / \partial s \partial t$ are 
normally distributed as $N(0, dtds )$; (2) for $t \in (0, 1)$, $\partial^2 G(t, t) / \partial t^2$ are normally distributed as $N(0, \frac{1}{2} (dt)^2)$; (3) for regions $B_i \subset B$, $i = 1, 2$, Cov$(G(B_1), G(B_2)) = |B_1 \cap B_2|$. For region $\widetilde{B} \subset B$, we define the integral

$$
G_{\widetilde{B}} = \int \int_{(s, t) \in \widetilde{B}} G(s, t) dsdt
$$

\noindent It is obvious that, as $n \rightarrow \infty$,

$$
\frac{|A_1|}{n^2} \rightarrow \frac{1}{2}t^2, \quad 
\frac{|A_1|}{|A_1 \cup R_t|} \rightarrow \frac{t}{2-t}, \quad 
\frac{|A_1 \cup R_t| }{|A|} \rightarrow t(2-t).
$$

\noindent Then by Donsker's theorem:

$$
\frac{S_{A_1}}{\sqrt{n^2/2}} \rightarrow G_{\widetilde{A}_1}, \quad 
\frac{S_{A_1 \cup R_t}}{\sqrt{n^2/2}} \rightarrow G_{\widetilde{A}_1 \cup \widetilde{R}_t}, \quad 
\frac{S_{A}}{\sqrt{n^2/2}} \rightarrow G_{\widetilde{A}}.
$$

\noindent Hence, 

$$
Z_m \rightarrow g_t, \quad 
\widetilde{Z} \rightarrow g_\delta.
$$

\noindent for $m_0/n \rightarrow \delta > 0$.

\hfill \(\Box\) \newline

\section*{The simulated distribution of $\widetilde{Z}$}

\begin{figure}[H]
\begin{center}
\includegraphics[scale=0.45]{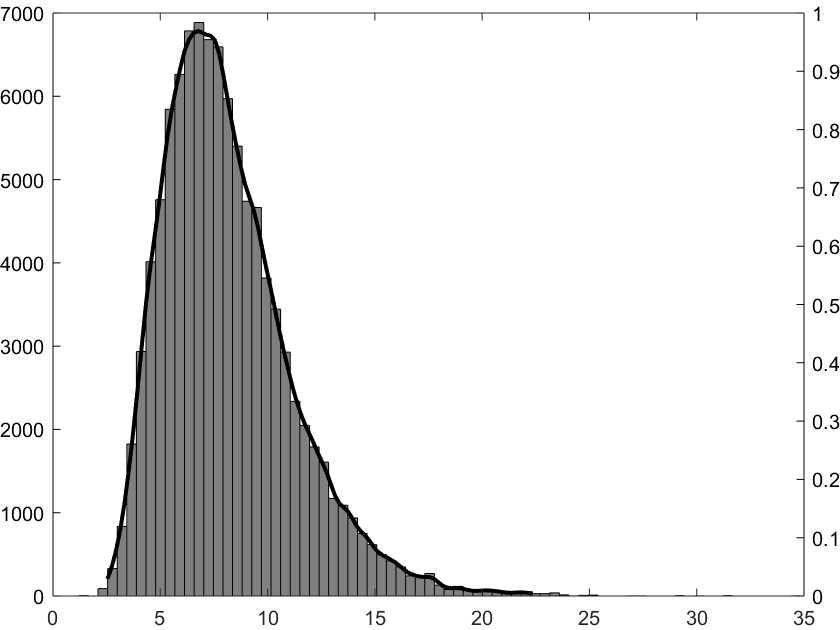}\par
\textbf{Fig S1.} Historgram and density function of simulated $\widetilde{Z}$ with $n=10^5$ and $m_0=n \cdot 2.5\%$.
\newline
\label{suppfig1}
\end{center}
\end{figure}

\section*{Theoretical and practical time complexity}

In best case, we have at most $r = log_2 n$ iterations. The time 
complexity is as following:

$$
\big(2\times \frac{n^2+n}{2}+4n\big) + \big(2^2\times \frac{(n/2)^2+n/2}{2}+4n\big) + 
... + \big(2^r\times \frac{(n/2^{r-1})^2+n/2^{r-1}}{2}+4n\big)
$$
$$
= n^2 \big( 1+\frac{1}{2}+\frac{1}{4}+... \big) + 5nr = O(n^2)
$$

In the worst case, one of the sub-matrix we divide is smallest possible 
with size $\xi$ in each iteration. We have at most $r = n/\xi$ iterations.

$$
\big(2\times \frac{n^2+n}{2}+4n\big) + \big(2\times \frac{(n-\xi)^2+(n-\xi)}{2}+
4(n-\xi)\big) + ... + \big(2\times \frac{(2\xi)^2+2\xi}{2}+4\times2\xi\big)
$$
$$
= \big(n^2 + (n-\xi)^2 + ... + (2\xi)^2\big) + 5\big(n+(n-\xi)+ ... + 2\xi\big) 
= O(n^3)
$$

We further tested HiCKey on interaction matrices of chr1 of GM12878 cell line.
The following Fig shows the real performance of running time and memory usage.

\begin{figure}[H]
\centering
\includegraphics[width=0.96\textwidth]{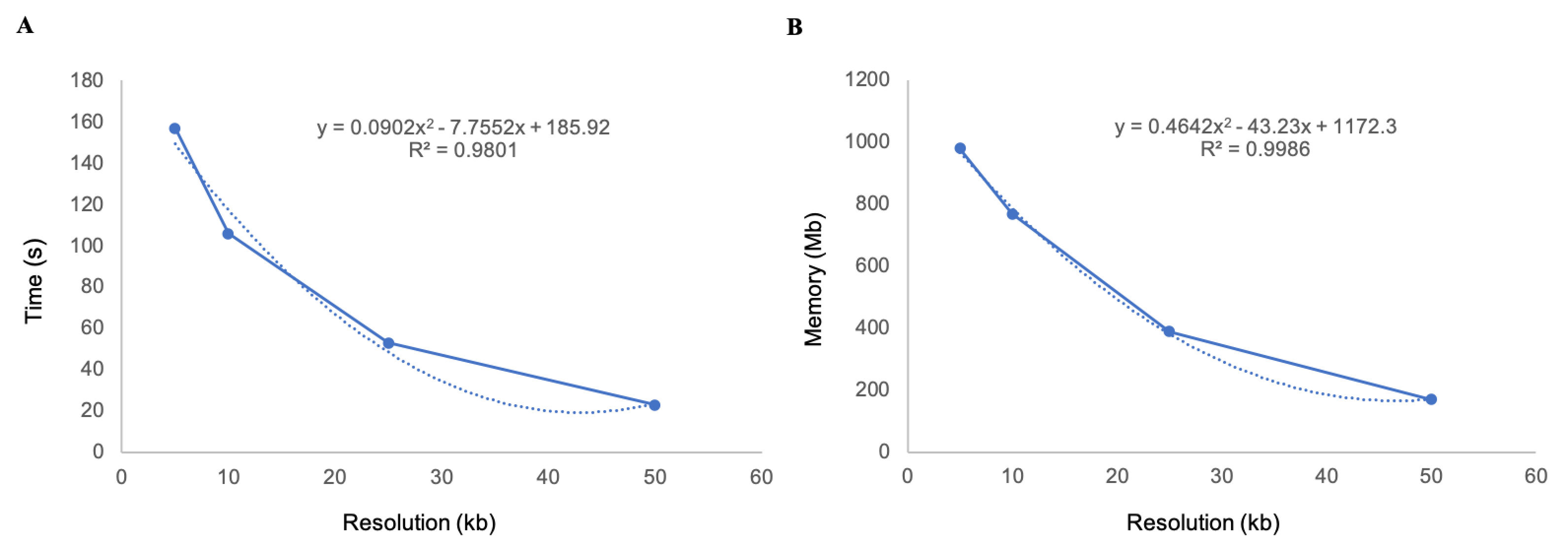}\par
\textbf{Fig S2. The practical memory usage and running time.} (\textbf{A}) Running time and resolution. (\textbf{B}) Memory and resolution. Both of the polynomial functions were estimated by using excel software. 
\label{suppfig2}
\end{figure}

\section*{Performance of five methods on simulation data (Sim1) and (Sim2)}

\begin{table}[H]
\textbf{Table S1} Performance of change-point estimation in (Sim1) at different noise levels. The numbers in each cell show the mean and standard deviation.

\centering
\label{tableS1}
\resizebox{0.75\textwidth}{!}{%
\begin{tabular}{c|c|c|c|c|c} \hline
\multicolumn{2}{c|}{} & \multicolumn{1}{c|}{$4\%$} & \multicolumn{1}{c|}{$8\%$} & \multicolumn{1}{c|}{$12\%$} & \multicolumn{1}{c}{$16\%$} \\ \hline
\multirow{3}{*}{HiCKey} & $\widehat{K}-K$ & -0.2 (0.20) & 1.8 (0.37) & 19.4 (3.22) & 82.6 (2.11) \\
& TPR & .9988 (.0012) & .9930 (.0022) & .9812 (.0039) & .9459 (.0087) \\
& FDR & .0000 (0) & .0173 (.0031) & .1176 (.0163) & .3618 (.0086) \\ \hline
\multirow{3}{*}{IC-Finder} & $\widehat{K}-K$ & 207.4 (24.11) & 113.8 (10.47) & 98.6 (9.24) & 104.0 (4.95) \\
& TPR & .5593 (.0335) & .8430 (.0461) & .8779 (.0247) & .8930 (.0106) \\
& FDR & .7441 (.0331) & .4902 (.0364) & .4395 (.0202) & .4412 (.0154) \\ \hline
\multirow{3}{*}{HiCSeg} & $\widehat{K}-K$ & -2.2 (.84) & -29.2 (10.16) & -99.2 (21.40) & -150.2 (3.27) \\
& TPR & .9872 (.0049) & .8290 (.0572) & .4232 (.1245) & .1244 (.0195) \\
& FDR & .0000 (.0000) & .0013 (.0029) & .0000 (.0000) & .0182 (.0407) \\ \hline
\multirow{3}{*}{TADTree} & $\widehat{K}-K$ & 397.6 (79.01) & 230 (63.84) & 145.8 (62.24) & 67.4 (34.64) \\
& TPR & .7791 (.0323) & .7826 (.0432) & .8035 (.0317) & .8465 (.0252) \\
& FDR & .8227 (.0348) & .7347 (.0578) & .6559 (.0718) & .5311 (.0685) \\ \hline
\multirow{3}{*}{3DNetMod} & $\widehat{K}-K$ & 519.40 (88.33) & 558.80 (95.99) & 424.60 (71.32) & 349.00 (30.82) \\
& TPR & .9205 (.0350) & .9345 (.0291) & .8947 (.0420) & .8491 (.0208) \\
& FDR & .7695 (.0249) & .7788 (.0217) & .7413 (.0197) & .7200 (.0169) \\ \hline
\end{tabular} }
\end{table}

\begin{table}[H]
\textbf{Table S2} Performance of change-point estimation in (Sim2) at different noise levels. The numbers in each cell show the mean and standard deviation.
\centering
\label{tableS2}
\resizebox{0.75\textwidth}{!}{%
\begin{tabular}{c|c|c|c|c|c} \hline
\multicolumn{2}{c|}{} & \multicolumn{1}{c|}{$4\%$} & \multicolumn{1}{c|}{$8\%$} & \multicolumn{1}{c|}{$12\%$} & \multicolumn{1}{c}{$16\%$} \\ \hline
\multirow{2}{*}{HiCKey} & TPR & .8450 (.0070) & .7570 (.0090) & .6568 (.0044) & .5289 (.0118) \\
& FDR & .0590 (.0050) & .0867 (.0058) & .1343 (.0043) & .1928 (.0066) \\ \hline
\multirow{2}{*}{IC-Finder} & TPR & .5508 (.0168) & .7010 (.0184) & .6858 (.0384) & .6533 (.0262) \\
& FDR & .5934 (.0266) & .3111 (.0208) & .2674 (.0348) & .2860 (.0430) \\ \hline
\multirow{2}{*}{HiCSeg} & TPR & .5071 (.0199) & .0990 (.0036) & .0350 (.0021) & .0258 (.0033) \\
& FDR & .0000 (.0000) & .0144 (.0211) & .0268 (.0368) & .0400 (.0894) \\ \hline
\multirow{2}{*}{TADTree} & TPR & .6990 (.0166) & .7071 (.0218) & .6756 (.0222) & .5467 (.1112) \\
& FDR & .5334 (.0372) & .4938 (.0245) & .4494 (.0377) & .4268 (.0394) \\ \hline
\multirow{2}{*}{3DNetMod} & TPR & .6015 (.0227) & .4748 (.0370) & .3522 (.0219) & .2738 (.0186) \\
& FDR & .5879 (.0186) & .5228 (.0138) & .5077 (.0228) & .5701 (.0065) \\ \hline
\end{tabular} }
\newline{}
\newline{}
\end{table}

\section*{The statistical results of hierarchical orders of 7 cell lines}

\begin{table}[H]
\textbf{Table S3} Hierarchical levels of TADs. The numbers in each cell show the TAD numbers and the percentage of total TAD.
\centering
\label{tableS3}
\resizebox{0.75\textwidth}{!}{%
\begin{tabular}{l|c|c|c|c|c} \hline
Cell line & $\#$TADs & order 1 & order 2 & order 3 & $\geq$order 4 \\ \hline
GM12878 & 8586 & 7743 (90.18$\%$) & 687 (8.00$\%$) & 96 (1.12$\%$) & 14 (0.16$\%$) \\ \hline
HMEC      & 8200 & 7072 (86.24$\%$) & 869 (10.60$\%$) & 183 (2.23$\%$) & 31 (0.38$\%$) \\ \hline
HUVEC    & 8903 & 8053 (90.45$\%$) & 690 (7.75$\%$) & 98 (1.10$\%$) & 16 (0.18$\%$) \\ \hline
IMR90     & 9043 & 8202 (90.70$\%$) & 689 (7.62$\%$) & 94 (1.04$\%$) & 12 (0.13$\%$) \\ \hline
K562       & 8801 & 8046 (91.42$\%$) & 618 (7.02$\%$) & 87 (0.99$\%$) & 4 (0.05$\%$) \\ \hline
KBM7      & 6246 & 5402 (86.49$\%$) & 647 (10.36$\%$) & 131 (2.10$\%$) & 21 (0.34$\%$) \\ \hline
NHEK      & 7726 & 6935 (89.76$\%$) & 652 (8.44$\%$) & 87 (1.13$\%$) & 6 (0.08$\%$) \\ \hline
\end{tabular} }
\newline{}
\newline{}
\end{table}

\section*{TAD boundary Enrichment in active chromosomal regions of 6 cell lines}

\begin{table}[H]
\textbf{Table S4} Estimated numbers of bins and boundaries in active and repressive regions by HiCkey. The enrichments of TAD boundaries (p-value) is calculated by using one-sided fisher-exact test. 
\centering
\label{tableS4}
\resizebox{0.75\textwidth}{!}{%
\begin{tabular}{l|l|c|c|r}
\hline
Cell Line                & State      & Bins & Boundaries & $p$-value  \\ \hline
\multirow{2}{*}{GM12878} & active     & 29293          & 2490    & \multirow{2}{*}{5.15e-18} \\ \cline{2-4}
                         & repressive & 81220          & 5553 \\ \hline
\multirow{2}{*}{HESC}    & active     & 55345          & 4540                 & \multirow{2}{*}{7.92e-4} \\ \cline{2-4}
                         & repressive & 13010          & 950 \\ \hline
\multirow{2}{*}{HUVEC}   & active     & 54380          & 4420                 & \multirow{2}{*}{4.48e-12} \\ \cline{2-4}
                         & repressive & 53832          & 3739 \\ \hline
\multirow{2}{*}{IMR90}   & active     & 64220          & 5494                 & \multirow{2}{*}{2.98e-58} \\ \cline{2-4}
                         & repressive & 46400          & 2706 \\ \hline
\multirow{2}{*}{K562}    & active     & 20960          & 1781                 & \multirow{2}{*}{3.50e-9} \\ \cline{2-4}
                         & repressive & 89932          & 6488 \\ \hline
\multirow{2}{*}{KBM7}    & active     & 34269          & 2090                 & \multirow{2}{*}{4.83e-15} \\ \cline{2-4}
                         & repressive & 75936          & 3716 \\ \hline
\end{tabular} }
\end{table}

\section*{Robustness of HiCKey against first iteration}

\begin{table}[H]
\textbf{Table S5} Robustness against first iteration for Sim1.
\centering
\resizebox{0.9\textwidth}{!}{%
\begin{tabular}{l|l|l|l|ll}
\hline
                         & 4\%        & 8\%        & 12\%      & 16\%      &  \\ \hline
$\hat{K}-K$ & 0.06(0.30)   & 0.02(0.32)   & 0.72(1.04)   & -1.05(1.33)   &  \\ \hline
TPR & .9977(.0035) & .9846(.0074) & .9761(.0129) & .9745(.0107) \\
\hline
\end{tabular}}
\end{table}

\begin{table}[H]
\textbf{Table S6} Robustness against first iteration for Sim2.
\centering
\resizebox{0.9\textwidth}{!}{%
\begin{tabular}{l|l|l|l|ll}
\hline
               & 4\%       & 8\%       & 12\%       & 16\%      &  \\ \hline
$\hat{K}-K$ & 0.58(0.86)   & 0.27(0.95)   & -0.93(1.92)    & -0.87(2.74)   &  \\ \hline
TPR & .9950(.0037) & .9804(.0079) & .93825(.0153) & .9073(.0303) & \\
\hline
\end{tabular}}
\end{table}

\begin{table}[H]
\textbf{Table S7} Robustness against first iteration for HESC.
\centering
\begin{tabular}{l|l|l}
\hline
chr            & $\hat{K}-K$  & TPR \\ \hline
chr1  & 0.42(1.66) & .96(.01) \\ \hline
chr2  & 0.74(2.10) & .94(.04) \\ \hline
chr3  & -0.66(1.50) & .96(.02) \\ \hline
chr4  & -0.68(1.88) & .94(.02) \\ \hline
chr5  & 1.21(1.47) & .96(.02) \\ \hline
chr6  & 1.91(2.51) & .94(.04) \\ \hline
chr7  & -1.35(1.51) & .96(.02) \\ \hline
chr8  & -0.24(1.11) & .96(.03) \\ \hline
chr9  & -0.85(2.05) & .97(.02) \\ \hline
chr10 & 1.33(2.00) & .95(.03) \\ \hline
chr11 & 0.06(1.16) & .95(.03) \\ \hline
chr12 & -0.38(1.77) & .95(.02) \\ \hline
chr13 & 0.14(1.80) & .93(.04) \\ \hline
chr14 & 0.13(1.47) & .95(.03) \\ \hline
chr15 & 1.65(1.27) & .95(.03) \\ \hline
chr16 & -0.08(1.09) & .97(.02) \\ \hline
chr17 & 0.10(1.02) & .95(.02) \\ \hline
chr18 & -0.53(1.18) & .93(.04) \\ \hline
chr19 & 0.21(1.27) & .90(.03) \\ \hline
chr20 & 0.02(0.82) & .94(.03) \\ \hline
chr21  & -0.26(1.11)  & .92(.06) \\ \hline
chr22 & -0.22(1.05)  & .95(.04) \\ \hline
chr23 & -1.52(2.77) & .94(.04) \\
\hline
\end{tabular}
\end{table}

\begin{table}[H]
\textbf{Table S8} Robustness against first iteration for IMR90.
\centering
\begin{tabular}{l|l|l}
\hline
chr            & $\hat{K}-K$  & TPR \\ \hline
chr1  & 0.39(1.64) & .97(.01) \\ \hline
chr2  & 0.16(1.68) & .95(.02) \\ \hline
chr3  & -1.44(1.63) & .95(.02) \\ \hline
chr4  & -0.28(1.81) & .95(.02) \\ \hline
chr5  & 0.68(1.76) & .96(.02) \\ \hline
chr6  & -2.09(1.63) & .94(.03) \\ \hline
chr7  & -0.72(1.11) & .97(.01) \\ \hline
chr8  & 0.51(1.19) & .96(.02) \\ \hline
chr9  & -0.49(1.20) & .96(.02) \\ \hline
chr10 & -0.25(1.21) & .96(.02) \\ \hline
chr11 & 0.01(1.22) & .97(.02) \\ \hline
chr12 & 0.30(1.27) & .96(.02) \\ \hline
chr13 & 1.26(1.48) & .93(.04) \\ \hline
chr14 & 1.45(1.34) & .94(.04) \\ \hline
chr15 & -0.37(1.35) & .97(.02) \\ \hline
chr16 & -0.07(1.59) & .96(.02) \\ \hline
chr17 & -2.50(1.45) & .93(.04) \\ \hline
chr18 & -0.04(1.03) & .96(.02) \\ \hline
chr19 & -0.33(1.06) & .92(.04) \\ \hline
chr20 & -0.36(0.95) & .94(.03) \\ \hline
chr21 & -0.18(0.93) & .91(.05) \\ \hline
chr22 & -0.90(1.55) & .94(.04) \\ \hline
chr23 & -0.19(1.70) & .96(.02) \\
\hline
\end{tabular}
\end{table}

\section*{Robustness of HiCKey against different distributions of HiC data}

\begin{table}[H]
\textbf{Table S9} Robustness against different distributions.
\centering
\begin{tabular}{l|l|l|l|l}
\hline
Setting                                    & noise      & $\hat{K}-K$      & TPR         & FDR         \\ \hline
\multirow{4}{*}{(S1)} & 0\%  & 0.15(0.02) & 1(0)        & .0046(4e-4) \\
                                           & 5\%  & 0.13(0.01) & 1(0)        & .0041(4e-4) \\
                                           & 10\% & 0.15(0.01) & 1(0)        & .0047(4e-4) \\
                                           & 15\% & 0.13(0.01) & .9998(9e-5) & .0043(4e-4) \\ \hline
\multirow{4}{*}{(S2)} & 0\%     & 0.14(0.01) & .9999(7e-5) & .0045(4e-4) \\
                                           & 5\%  & 0.16(0.01) & .9998(9e-5) & .0050(4e-4) \\
                                           & 10\%   & 0.15(0.01) & .9998(7e-5) & .0048(4e-4) \\
                                           & 15\%  & 0.14(0.01) & .9998(7e-5) & .0055(4e-4) \\
\hline
\end{tabular}
\end{table}

\medskip

\bibliographystyle{vancouver}  
\bibliography{mybib.bib}      

\end{document}